\documentstyle[psfig]{l-aa}

\newcommand {\msol}{\mbox{M$_{\odot}$}}
\newcommand {\mdust} {\hbox{$M_{\rm dust}$}}
\newcommand{\mhtwo} {\hbox{$M_{{\rm H}_2}$}}
\newcommand{\mhi}   {\hbox{$M_{\rm HI}$}}
\newcommand{\lo}    {\hbox{${\rm L}_{\odot}$}}
\newcommand{\lsol}    {\hbox{${\rm L}_{\odot}$}}
\newcommand{\lb}    {\hbox{$L_{\rm B}$}}
\newcommand{\lfir}  {\hbox{$L_{\rm FIR}$}}

\newcommand{\ssix}  {\hbox{$S_{60\mu \rm m}$}}           
\newcommand{\shun}  {\hbox{$S_{100\mu \rm m}$}}          

\newcommand{\ltwogiga}{$L_{2.4GHz}$}
\newcommand{\kms}   {\hbox{${\rm km\,s}^{-1}$}}
\newcommand{\kmsmpc}{\hbox{${\rm km\,s}^{-1}{\rm Mpc}^{-1}$}}

\newcommand{\tdust} {\hbox{$T_{\rm d}$}}
\newcommand{\tastar}{\hbox{$T_{\rm A}^{*}$}~}
\newcommand{\Kkms}  {\hbox{${\rm K\,km\,s}^{-1}$}}

\newcommand{\hI}    {\hbox{${\rm HI}$}}

\newcommand{\chitwo}   {\hbox{$\chi^2$}}
\newcommand {\CO}[2]{\mbox{$^{12}$CO(#1$\rightarrow$#2)}}
\newcommand{\hi}    {\hbox{${\rm HI}$}}			
\newcommand{\htwo}  {\hbox{${\rm H}_2$}}                
\begin{document}
\thesaurus{ 11 (11.07.1;  
	    11.09.2;  
	    11.19.3;  
            11.19.7;  
	           09.13.2;  
	           13.19.1)   
	         }
   \title{Molecular gas in galaxies of Hickson compact groups}


   \author{S. Leon \inst{1} \and F. Combes \inst{1} \and T. K. Menon \inst{2}}

   \offprints{S. Leon}

   \institute{DEMIRM, Observatoire de Paris, 61 Av. de l'Observatoire, 
F-75 014, Paris, France \and Dep. of Physics \& Astronomy, Univ. of 
British Columbia, 2219 Main Mall, Vancouver B.C., V6T 1Z4, Canada}

   \date{}
   \maketitle

   \begin{abstract}

We have observed 70 galaxies belonging to 45 Hickson compact groups in
the \CO{1}{0} and \CO{2}{1} lines, in order to determine their
molecular content. We detected 57 galaxies, corresponding to a
detection rate of 81\%. We compare the gas content relative to blue and
\lfir\ luminosities of galaxies in compact groups with respect to other
samples in the literature, including various environments and
morphological types. We find that there is some hint of  
enhanced \mhtwo/\lb\ and \mdust/\lb\ ratios in the galaxies from compact group
 with respect to our control
sample, especially for the most compact groups, suggesting that tidal
interactions can drive the gas component inwards, by removing its
angular momentum, and concentrating it in the dense central regions,
where it is easily detected. The molecular gas content in compact group
galaxies is similar to that in pairs and starburst samples.  However,
the total \lfir\ luminosity of HCGs is quite similar to that of the
control sample, and therefore the star formation efficiency appears
lower than in the control galaxies. However this assumes that the FIR
spatial distributions are similar in both samples which is not  the
case at radio frequencies.  Higher spatial resolution FIR data are
needed to make a valid comparison.  Given their short dynamical
friction time-scale, it is possible that some of these systems are in
the final stage before merging, leading to ultra-luminous starburst
phases.  We also find for all galaxy samples that the \htwo\ content
(derived from CO luminosity and normalised to blue luminosity) is
strongly correlated to the \lfir\ luminosity, while the total gas
content (\htwo+\hi) is not.

   \end{abstract}

\section{Introduction}
Galaxies are gregarious systems, most of them are gathered in groups or
clusters, while only 30\% are isolated and 10\% are binaries in the
field.  Nevertheless compact groups (CG) are quite rare and according
to Hickson's classification (Hickson, 1982) only 0.1 \% of galaxies
belong to CGs.  Criteria of population (initially four galaxies in the
group), isolation (dynamically independent systems) and compactness
(separation between galaxies comparable to the sizes of the galaxies)
are chosen by Hickson to build his catalog.  With these criteria around
one hundred CGs were found on the Palomar Observatory Sky Survey red
prints.

 Compact groups are ideal sites to study the influence of strong
dynamical evolution due to environment on molecular cloud formation and
star formation efficiency. They appear in projection as the densest
galaxy systems known, even denser than the cores of rich clusters, and
they show multiple signs of interactions.  Due to their high density,
and relatively small velocity dispersion, these systems are unstable
with regard to merging instability. The dynamical friction time-scale
is of the order of 2.10$^8$ yrs, and N-body simulations predict their
rapid evolution towards a single elliptical massive galaxy (e.g. Barnes
1989). The existence of many such compact groups is therefore a puzzle,
and the physical reality of HCG has been questioned (e.g. Mamon 1986,
1987); but evidence of galaxy-galaxy interactions in those groups,
either morphologic (Hickson 1990; Mendes de Oliveira 1992), or
kinematic (Rubin et al. 1991), speaks in favour of their reality.
Latest spectroscopic observations showed that 92 of the original 100
groups have at least three galaxies with recession velocities within
1000 \kms of each other (Hickson et al. 1992). The presence of hot 
intergalactic gas,
detected by X-ray emission centered on some HCGs, is a further
confirmation of the existence of these compact groups (Pildis et al.
1995, Ponman et al. 1996).\\

Most of galaxies that belong to groups are in fact in loose groups of
10-30 galaxies and about 50\% of all galaxies belong to loose groups.
But loose groups are in their great majority un-bound and un-virialised
(Gourgoulhon et al. 1992) while  their true dynamical state is
ambiguous (expanding, collapsing, transient).  Clusters of galaxies are
more near equilibrium, specially in their centers (about 10\% of all
galaxies belong to clusters). However, the depth of their potential
well leads to high relative velocities between galaxies that reduce the
efficiency of galaxy-galaxy encounters. The influence of environment is
revealed by the high proportion of ellipticals and lenticulars, and by
the HI gas deficiency of spirals (Dressler 1984, Cayatte et al. 1991).
This gas deficiency can be explained by ram-pressure as well as tidal
interactions (Combes et al. 1988). No molecular gas deficiency has been
detected, either in Virgo (Kenney \& Young 1988), or in Coma (Casoli et
al. 1991), which suggests that the inner parts of the galaxies are not
affected by their environment, since the CO emission essentially comes
from the galaxy central regions. However, there could be two
compensating effects at play here: the enhancement of CO emission in
interacting galaxies (cf Braine et Combes 1993, Combes et al. 1994),
and the outer gas stripping, stopping the gas fueling of galaxies.\\

 In compact groups, some HI deficiency has also been reported (Williams
\& Rood 1987), but no CO emission deficiency, according to a first
study by Boselli et al (1996) with the SEST telescope.  It is further
interesting to investigate whether HCGs are actually sampling the
highest densities of galaxies in the Universe.  It has been claimed
that, even if the CGs are real, we are not sure of their high density,
since they could correspond to loose groups with only a high {\it
projected} density through chance alignment of filaments along the line
of sight (e.g. Mamon 1992). But no loose groups are observed in HCG
neighborhood in the majority (67\%) of cases (Rood \& Williams 1989).
Hickson (1982) found that the groups contain fewer spirals than a
comparable sample of field galaxies. The spiral fraction decreases from
60\% in the least compact groups to 20\% in the most compact. There is
also a deficiency of faint galaxies with respect to rich clusters and
field. This apparent deficiency is more severe in groups with
elliptical first-ranked galaxies.  Radio properties of compact groups
have been studied by Menon \& Hickson (1985) and Menon (1991, 1995).
Although the far-infrared and radio luminosities are still highly
correlated as for field galaxies, the total radio emission from HCG
spirals is relatively lower by a factor 2 in compact group galaxies
while the nuclear radio emission is enhanced by a factor of about 10
compared to isolated galaxies. The results suggest a scenario in which
interactions among group galaxies produce inflow of gas towards the
centers, elevating the star formation there, and consequently the radio
and far-infrared emissions. But at the same time the removal of gas and
magnetic fields from the extended disks of the galaxies results in a
decrease of total radio-emission.  Williams \& Rood (1987) have
observed 51 of the 100 Hickson groups in the HI line, and detected 34
of them. They find that on average a Hickson compact group contains
half as much neutral hydrogen as a loose group with a similar
distribution of galaxy luminosities and morphological types.  This
result supports the reality of compact groups as independent dynamical
systems and not transient or projected configurations in loose groups.
The recent ROSAT survey of HCGs by Ponman et al (1996) also confirms
that the groups are intrinsically compact, and not the projection of
loose groups. They infer that more than 75\% of the HCGs possess hot
intragroup gas.\\

 We present here a large CO survey of Hickson group galaxies with the
IRAM 30m telescope, and compare the relative gas content and star
formation efficiency of CG galaxies with other samples belonging to
widely different environments.  After describing the observations in
section 2, the sample and the data in section 3, we discuss the main
conclusions and  possible interpretations in section 4.

\section{Observations}

Observations were carried out on 4-10 September 1995 with the 30 meter
radiotelescope of the Instituto de Radio Astronomia Milimetrica (IRAM)
in Pico Veleta, near  Granada in Spain. Single-sideband SIS receivers
were tuned for the \CO{1}{0} and \CO{2}{1} transitions at respectively
115 and 230 GHz. Weather conditions were excellent during the run with
typical effective system temperatures of 300-400 K (\tastar scale ) at
115 GHz and  500-800 K (\tastar  scale) at 230 GHz.\\ For each line a
512 $\times $1 MHz channel filter bank is used with a velocity
resolution of 2.6 \kms smoothed for each spectrum to 10.4 or 20.8 \kms
according the quality of each spectrum. At 115 GHz and 230 GHz we
assume a HPBW of 22$^{\prime\prime}$ and 11$^{\prime \prime}$
respectively.  Pointing was done frequently on continuum  sources with
corrections of the offsets up to 8$^{\prime \prime}$, providing an
accuracy of 3$^{ \prime \prime}$ (Greve et al, 1996).\\

The temperature-scale calibration was checked on the sources W3OH, ORIA
and IRC+10216 (Mauersberger et al. 1989); except for a transient
problem for the 3mm calibration, it remained in a reasonable range
providing at least a 20\% calibration accuracy. We use the wobbler with
a switch cycle of 4 seconds and a beam throw of 90-240$^{\prime
\prime}$  avoiding off position on another galaxy of the same compact
group.  Each 12 minutes a chopper wheel calibration was performed on a
load at ambient temperature and on a cold load (77 K). The line
temperatures are expressed in the \tastar scale, antenna temperature
corrected for atmospheric attenuation and rear sidelobes. Baselines
were flat allowing us to subtract only linear polynomials out of the
spectra.\\

\section{Results} 

\subsection{The observed sample} 

Our observed sample is composed of 70 galaxies towards 45 compact
groups, taken from the catalog of Hickson Compact Groups (HCG, Hickson
1982) .  We discarded afterwards 4 galaxies (11a,19b,73a,78a) which
appear not to belong to Compact Groups (Hickson, 1992). The galaxies,
mostly  spirals, are selected for their radio continuum (Menon 1995)
and IRAS (Hickson et al. 1989) detections. All the targets are northern
sources with $\delta > 0^\circ$. The redshift range from 1200 \kms\  up
to 18500 \kms\ (27b).

\begin{figure}
\psfig{width=9.5cm,height=9cm,file=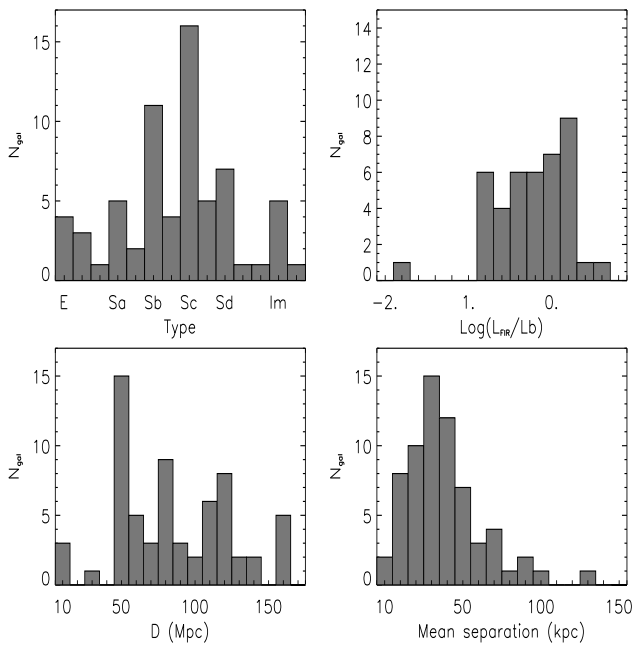}
\caption{Statistical properties of our CO sample.}
\label{stat_sample}
\end{figure}

If we use H$_0$=75 \kmsmpc\ for the Hubble constant, as adopted in this
paper, the mean distance of the sample is 95 Mpc with a standard
deviation of 45 Mpc.  We present in Fig. \ref{stat_sample} the
statistical distribution of our sample for distance, type, far infrared
(\lfir) luminosity and median projected separation. \\

In Table \ref{tab_sample_1} we display the main properties of the
sample. The column headings are the following: name is taken from
Hickson classification, type  is  taken from Hickson et al. (1989), D
is the distance computed with a correction for the galactic rotation
using a solar galactic velocity of rotation of 250 \kms, R is the
median projected galaxy-galaxy separation, $D_B$ indicates the diameter
in arcsecs at $\mu_B$ = 24.5 mag.arcsec$^{-2}$, from Hickson et al.
(1989), blue luminosity L$_B$ is computed as follows
\lb=12.208-0.4B$^o_T$+log(1+z)+2log(D/Mpc), \lfir\ luminosity is
computed  using reprocessed IRAS data from Allam et al. (1996) and
following Hickson et al. (1989) derivation, \tdust\ is  a dust
temperature indicator using an emissivity dependence as $\lambda^{-1}$
in the 60-100 $\mu$m IRAS- range, S$_{20cm}$ is the total radio
continuum flux density from Menon (1995), neutral hydrogen content has
been found mainly from Williams \& Rood (1987), and some from
Huchtmeier \& Richter (1989).   
Concerning \lfir\ luminosity or HI mass,
we indicate in Table \ref{tab_sample_1} the whole group emission preceded by $\leq$,
 meaning that poor spatial resolution does not allow
a separation per galaxy.\\

\begin{table*}

\begin{tabular}{clcccccccccc}
\hline
\hline
Name &type &D & R & D$_B$ &Log(L$_B$) &Log(L$_{FIR}$)&T$_d$ & S$_{20\mbox{cm}}$
 & Log(M(HI)) &Log(M$_{dust}$)  \\
     &   & (Mpc)  & (kpc)  & ($\prime\prime$) &  &  & (K) & (mJy)  
& (\msol) & (\msol) \\   
\hline
2b  & cI    & 59.3   & 52.5  & 37.6  & 9.95  & 10.16  & 35.8  &       &$\leq$10.11  & 6.61\\
3c  & Sd    & 103.0  & 77.0  & 19.8  & 9.93  & 9.93   & 43.3  &	      &$<$10.15   & 5.94\\
4a  & Sc    & 112.3  & 57.0  & 67.5  & 10.86 & 10.95  & 34.6  &       &                   & 7.49\\
7a  & Sb    & 57.8   & 45.6  & 80.2  & 10.50 & 10.27  & 35.2  & 12.85 & $\leq$9.28 & 6.76\\ 
7c  & SBc   & 57.8   & 45.6  & 91.0  & 10.19 & 9.69   & 29.0  &       & 9.64   & 6.68\\ 
10a & SBb   & 66.6   & 92.9  & 93.6  & 10.76 & 9.01   &       &       & 9.90   &\\
10c & Sc    & 66.6   & 92.9  & 49.2  & 10.18 & 9.84   & 32.0  & 2.47  & 9.61   &6.58\\
11a$^{\dag}$& SBbc& 73.1& -  & 91.9  & 10.70 & 9.64   & 27.7  &       &                   &6.77\\
14b & E5    & 73.5   & 26.9  & 51.9  & 10.23 & $<$9.12  &       &       & $<$9.77      & \\
14c & Sbc   & 73.5   & 26.9  & 15.9  & 9.26  & $<$9.12  &       &       & $<$9.77      & \\
16a & SBab  & 52.5   & 44.6  & 77.1  & 10.50 & 10.53  & 29.6  & 33.18 & 9.48     & 7.47\\
16b & Sab   & 52.5   & 44.6  & 61.3  & 10.30 & $<$8.84  &       & 2.62  & $\leq$10.07 & \\
16c & Im    & 52.5   & 44.6  & 58.9  & 10.36 & 10.69  & 33.8  & 78.29 & $\leq$10.07 & 7.28\\
16d & Im    & 52.5   & 44.6  & 63.5  & 10.24 & 10.74  & 33.9  & 30.91 & $\leq$10.07 & 7.33 \\ 
19b$^{\dag}$&Scd&56.1 & -    & 28.5  & 9.49  & 9.37   & 28.7  & 2.08  & $\leq$9.28        & 6.40 \\
21a & Sc    & 99.8   & 134.9 & 44.9  & 10.51 & 10.36  & 30.8  &       &           & 7.20\\ 
23b & SBc   & 64.0   & 66.1  & 48.7  & 10.00 & 10.09  & 31.2  & 8.01  & $\leq$10.04 & 6.87\\
23d & Sd    & 64.0   & 66.1  & 26.7  & 9.38  & 9.29  &       & 3.18  & $<$9.26       & \\ 
25a & SBc   & 84.6   & 47.9  & 51.1  & 10.47 & 9.93   & 32.6  & 4.70  & $\leq$10.09      & 6.62 \\
25c & Sb    & 145.0  & 47.9  & 23.9  & 10.35 & 10.44  & 34.0  &       &                  & 7.03 \\
27b & SBc   & 245.7  & 107.2 & 21.7  & 10.66 & 10.54 &  31.4&       &             & 7.32 \\
31a & Sdm   & 57.0   & 49.0  & 33.6  & 9.74  & $\leq$10.32 &  &       & 9.92       &  \\
31c & Im    & 57.0   & 49.0  & 74.5  & 10.68 & $\leq$10.32 &  &       & 9.85       & \\
33c & Sd    & 103.7  & 24.5  & 18.7  & 9.64  & 9.86  & &4.58 & $\leq$10.18 & \\ 
34b & Sd    & 121.8  & 15.5  & 22.2  & 9.71  & $\leq$10.36 &  & 6.03  & $<$10.00      & \\
37b & Sbc   & 88.5   & 28.8  & 40.8  & 10.26 & 9.85  & 28.9 & 1.55  & $<$8.90      & 6.85 \\
38b & SBd   & 115.1  & 58.9  & 38.4  & 10.38 & $\leq$10.55 &  &       & $\leq$9.70  & \\
40a & E3    & 86.7   & 15.1  & 65.3  & 10.66 & $\leq$10.09 &  &       & $<$9.70      & \\
40c & Sbc   & 86.7   & 15.1  & 36.5  & 9.98  & 10.07 & 29.4 & 6.03  & $<$9.70       & 7.03 \\
40d & SBa   & 86.7   & 15.1  & 41.0  & 10.23 & 9.56       &       & 6.39  & $<$9.70       &\\
40e & Sc    & 86.7   & 15.1  & 17.8  & 9.36  & 9.63 &  54.2&       & $<$9.70      &5.21\\
43a & Sb    & 129.8  & 58.9  & 27.0  & 10.34 & 10.18 & 25.4& 0.95& $<$10.16    & 7.56\\
43b & SBcd  & 129.8  & 58.9  & 25.8  & 10.32 & 10.12  & 26.7  & 1.06  & $<$10.16     & 7.34\\
44a & Sa    & 17.3   & 38.0  & 130.5 & 10.03 & 9.31   & 31.4  & 4.77  & 8.64      & 6.09\\
44c & SBc   & 17.3   & 38.0  & 91.5  & 9.62  & 8.94   & 32.2  & 2.45  & 8.41      & 5.66\\
\hline

\end{tabular}

\caption{General properties of our sample}
\label{tab_sample_1}
{$^{\dag}$not included in our final HCG sample}
\end{table*}


\addtocounter{table}{-1}
\begin{table*}
\begin{tabular}{clccccccccccc}
\hline
\hline
Name &type &D & R & D$_B$ &Log(L$_B$) &Log(L$_{FIR}$)&T$_d$ & S$_{20\mbox{cm}}$
 & Log(M(HI)) &Log(M$_{dust}$)  \\
     &   & (Mpc)  & (kpc)  & ($\prime\prime$) &  &  & (K) & (mJy)  
& (\msol) & (\msol) \\   
\hline
44d & Sd    & 17.3   & 38.0  & 70.3  & 9.40  & 8.65   & 39.0  & 2.95  & 8.88     & 4.90\\  
47a & SBb   & 125.3  & 36.3  & 40.3  & 10.52 & 10.27  & 32.1  & 12.25 & $<$9.78      & 7.00\\
49a & Scd   & 134.2  & 12.3  & 21.6  & 10.07 & $\leq$10.12 &  & 0.68  &            & \\
49b & Sd    & 134.2  & 12.3  & 15.9  & 9.90  & $\leq$10.12 &  & 1.70  &            & \\
55a & E0    & 212.0  & 19.1  & 23.1  & 10.64 & $\leq$10.58 &  &       &            & \\
57d & SBc   & 120.8  & 72.4  & 44.3  & 10.52 & $<$10.17 &       & 5.09  &             & \\
58a & Sb    & 81.5   & 89.1  & 58.2  & 10.56 & 10.59 & 34.3 & 20.71 & 9.75       & 7.14\\
59a & Sa    & 53.0   & 21.4  & 38.1  & 9.80  & 10.16   & 46.4  & 9.02  & 9.26       & 6.03\\
59d & Im    & 53.0   & 21.4  & 28.0  & 9.29  & $<$10.16   &   &     & $\leq$8.97  & \\
61c & Sbc   & 51.9   & 28.8  & 57.4  & 10.18 & 10.41  & 33.9  &       & $\leq$9.95  & 6.99\\
61d & S0    & 51.9   & 28.8  & 38.7  & 9.95  & $<$8.92  &       &       & $\leq$9.95  &\\
67b & Sc    & 96.8   & 49.0  & 48.1  & 10.58 & 10.28  & 29.7  & 6.70  & 10.20      & 7.22\\
67c & Scd   & 96.8   & 49.0  & 38.7  & 10.11 & $<$9.87  &       & 13.49 & 9.54        & \\
68c & SBbc  & 33.0   & 33.1  & 134.3 & 10.43 & 9.79   & 28.3  & 2.10  & 9.65        & 6.86\\
69a & Sc    & 117.9  & 30.2  & 39.4  & 10.33 & $<$9.48  &       & 3.13  & $\leq$10.15 & \\
69b & SBb   & 117.9  & 30.2  & 23.8  & 10.07 & 10.67  & 41.4  & 6.60  & $\leq$10.15 & 6.79\\
71b & Sb    & 120.9  & 50.1  & 22.6  & 10.37 & 10.60  & 35.0  &       & $<$10.27    & 7.11\\
73a$^{\dag}$ & Scd   & 76.9   & 100.0 & 76.3  & 10.62 & 9.32&       &       & 10.10             & \\
75b & Sb    & 167.3  & 37.2  & 14.8  & 10.65 & $<$9.89  &       & 3.10  &             & \\
75e & Sa    & 167.3  & 37.2  & 15.1  & 10.07 & $\leq$10.22 &  &       &            & \\
78a$^{\dag}$ & SBb   & 116.8  & -     & 51.1  & 10.56 & 10.36  & 31.6  &       & 10.29     & 7.12 \\
79a & E0    & 59.3   & 6.8   & 44.7  & 9.97  & $\leq$9.82 &   &       & $<$9.15       & \\
79c & S0    & 59.3   & 6.8   & 33.8  & 9.82  & $<$9.19  &       &       & $<$9.15      & \\
80a & Sd    & 126.4  & 25.1  & 25.4  & 10.46 & 10.84 & 33.9 & 20.62 &         &7.43\\
82c & Im    & 146.7  & 70.8  & 32.9  & 10.58 & 10.59  & 33.9  & 8.33  & $<$10.09      & 7.17\\
88a & Sb    & 82.3   & 67.6  & 57.3  & 10.72 & 10.01  & 24.8  & 0.90  & $<$10.02     & 7.46\\
89c & Scd   & 120.8  & 58.9  & 24.4  & 10.12 & $<$9.94  &       &       & $<$10.46     & \\
92c & SBa   & 89.0   & 28.2  & 83.2  & 10.73 & 10.15  & 26.0  & 24.74 & 9.90        & 7.46\\ 
93b & SBd   & 69.8   & 70.8  & 64.1  & 10.58 & 10.25  & 31.4  & 10.92 & 9.59       & 7.03\\
95b & Scd   & 160.9  & 30.2  & 25.1  & 10.44 & $\leq$10.69 &  & 3.86  &            & \\
95c & Sm    & 160.9  & 30.2  & 26.3  & 10.50 & $\leq$10.69 &  & 6.06  &            & \\
95d & Sc    & 160.9  & 30.2  & 16.9  & 10.12 & 	      &       &       &            & \\
96a & Sc    & 119.0  & 30.2  & 61.6  & 10.90 & 11.10  & 39.4  & 200.1 & $<$10.16     & 7.33\\
96c & Sa    & 119.0  & 30.2  & 19.8  & 10.04 & $<$9.88  &       & 4.80  & $<$10.16    & \\
100a & Sb   & 73.3  & 38.0  & 48.9  & 10.43 & 10.273 & 32.6  & 9.17  & 9.74       & 6.95\\
\hline

\end{tabular}
\caption{General properties of our sample (following)}
\label{tab_sample_2}
{$^{\dag}$not included in our final HCG sample}
\end{table*}

\subsection{Comparison samples}

We compare our results on Hickson compact groups with a 'control'
sample, gathering most of the CO data obtained in the literature until
now: about 200 galaxies observed by Young et al (1989, 1996) with the
FCRAO 14m antenna, by Solomon \& Sage (1988) with the FCRAO and NRAO
Kitt Peak 12m telescopes, Tinney et al (1990) and by Sage (1993) with
the NRAO antenna. This big control sample consists essentially of
nearby bright galaxies, and includes a wide range of environment
conditions, from isolated and field, to interacting systems.  The
mergers are also included, and correspond to the highest FIR
luminosities of the ensemble. From these observations, we have however
separated the Virgo galaxies, and added the results on Coma galaxies
(Casoli et al 1991), to build the sample 'Cluster'.  We also compare
HCG data to more specific samples, such as the isolated pairs from
Combes et al (1994), dwarfs from Sage et al (1992), Israel et al (1995)
and Leon et al. (1997), and ellipticals from Wiklind et al. (1995), and
from Sanders et al. (1991) for the starbursts. For this latter ensemble
we separate the pair galaxies to include them in the pair sample.
These are noted respectively 'pair', 'dwarf', 'elliptic' and
'starburst' in the various figures of the present work. We have summarized 
the size of the different samples in Table \ref{table_sample}.\\

\begin{table}

\begin{tabular}{ccc}
\hline
\hline
sample &  number of galaxies  & references \\
\hline
control	& 193 &	7,8,9,10\\
starburst & 73 & 5\\
pair	& 48 & 2,5\\
cluster	& 40 & 1,10\\ 
dwarf	& 25 & 3,4,6\\
elliptic & 18 & 11 \\
\hline
\end{tabular}
\begin{minipage}{6cm}
{\scriptsize
(1) Casoli et al. (1991), (2) Combes et al. (1994), (3) Israel et al. (1995), 
(4) Leon et al. (1997), (5) Sanders et al. (1991), (6) Sage et al.(1992), (7) Sage (1993),
(8) Solomon \& Sage (1988), (9) Tinney et al. (1990), (10) Young et al. (1989,1996),
 (11) Wiklind et al. (1995)
}
\end{minipage} 
\caption{Galaxy numbers in the comparison samples.}
\label{table_sample}
\end{table}

\subsection{CO data}

\begin{figure*}
\psfig{width=19cm,height=25cm,file=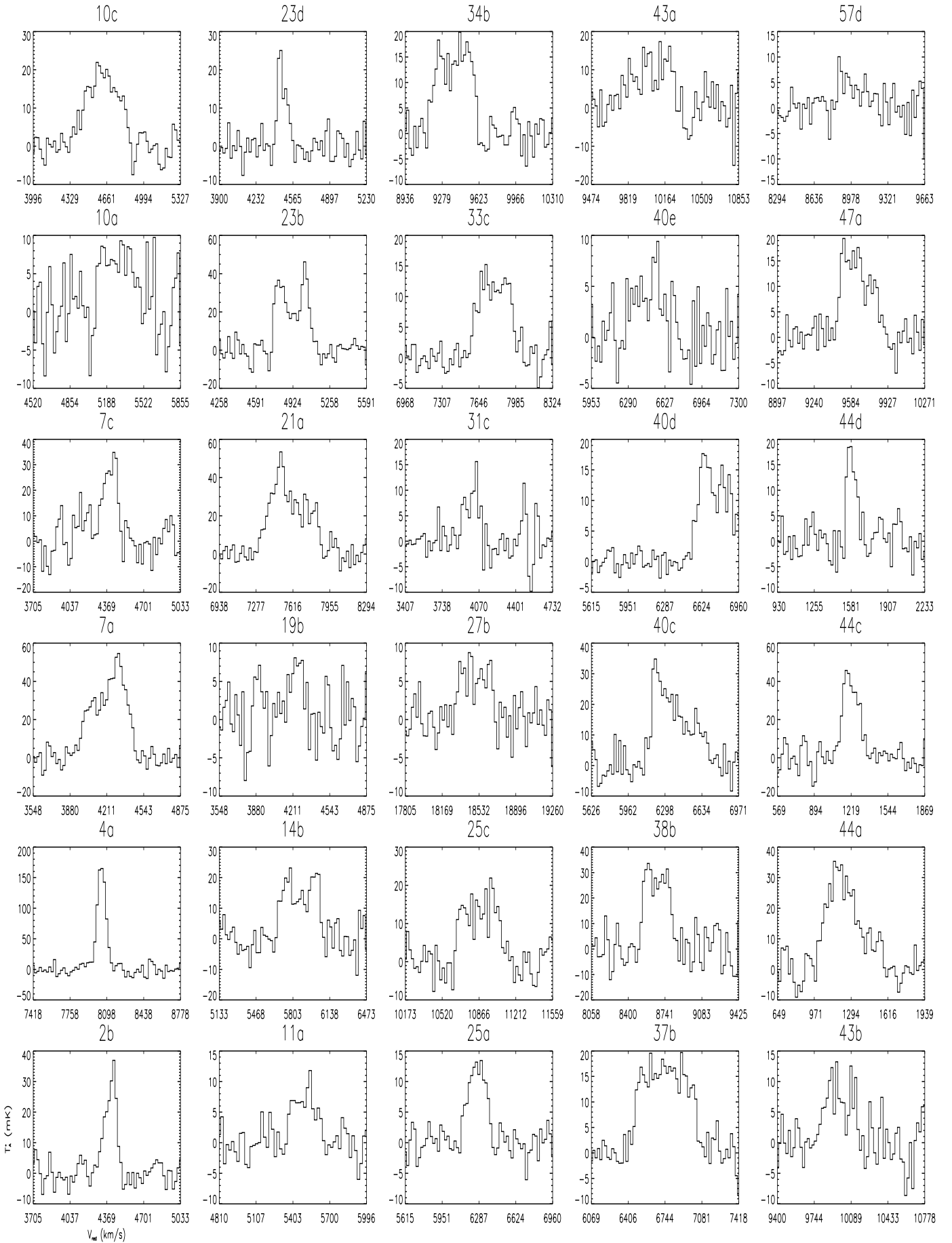}
\caption{Line profiles of the CO(1-0) emission from our HCG sample. The 
intensity scale is in units
of \tastar (mK). The x-axis is the redshift expressed in \kms.}
\label{fig_spectra}
\end{figure*} 

\addtocounter{figure}{-1}
\begin{figure*}
\psfig{width=19cm,height=25cm,file=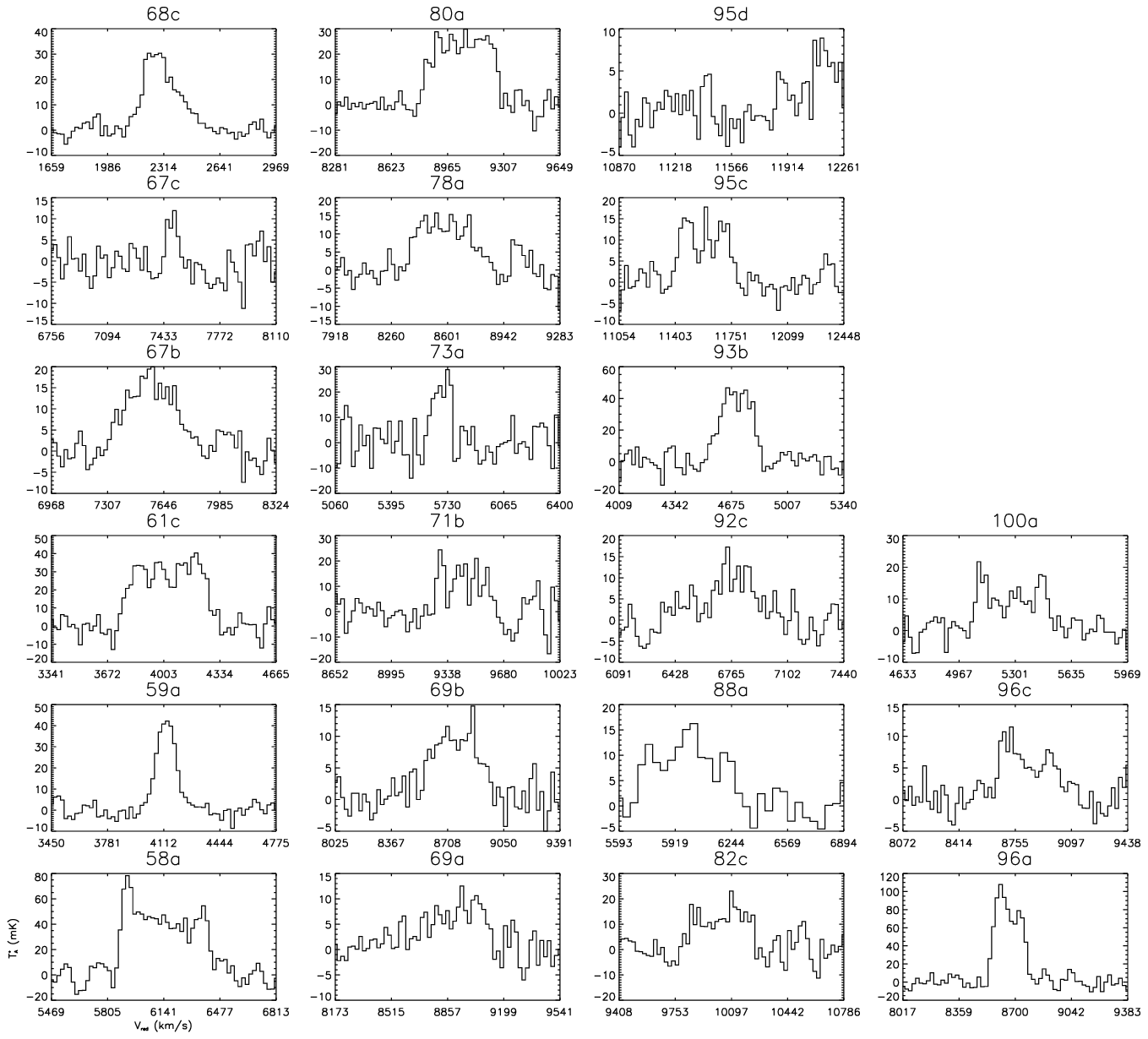}
\caption{continued}
\end{figure*}

We detected 57 (or 53 in our final HCG sample) galaxies, corresponding
to a 80 \% detection rate, with 2 detections in \CO{2}{1} (75e,89c) not
detected in \CO{1}{0}, probably due to dilution factor, given  the
small size of these two galaxies. Some galaxies were detected only in
the \CO{1}{0} line.  We reached typically an rms temperature level of
1-8 mK at a smoothed velocity resolution of 10.4-20.8 \kms according to
the spectra quality.  In Table \ref{tab_co_1} we present the results of
our CO observations:  I$_{CO}$ is the velocity-integrated temperature
$\int T_a^* dv$ for the \CO{1}{0} line, or integrated  CO intensity, in
the \tastar scale, $\delta$I$_ {CO}$ is the standard error on I$_{CO}$, 
$T_p$ is the peak antenna temperature, 
v$_{CO}$ is the intensity-weighted mean heliocentric velocity, FWHM is the 
full width half maximum of the  \CO{1}{0} spectra, I$_{CO(2-1)}$ is the 
\CO{2}{1} integrated intensity and M(H$_2$) is the molecular
gas mass. The spectra of galaxies detected through the CO(1-0) line are
displayed in Fig. \ref{fig_spectra}.\\
Upper limits of the CO intensities are computed at 3$\sigma$ in antenna
temperature scale, and with a line width $\delta_{CO}$ guessed from
other lines when available from \hi\, lines (Williams \& Rood 1987) , and 
taken as $\delta_{CO}$ = 200
\kms\ otherwise. In the case of 40d and 95d we deduced
I$_{CO(2-1)}$ and I$_{CO}$ intensities from the fit, because the emission was 
shifted at the edge of the band.  To derive H$_2$ molecular gas content the
standard H$_2$/CO conversion factor is adopted (Strong et al. 1988),
i.e:

\begin{equation}
N(H_2)=2.3\times10^{20}\int_{line} T_R dv  \mbox{(mol.cm$^{-2}$)}
\end{equation}
where $T_R$ is the radiation temperature. Following Gordon et al.
(1992) to include size source correction,  H$_2$ mass is derived using
the expression (see appendix A):

\begin{equation}
M(H_2)=5.86\times10^4D^2KI_{CO} (\msol)
\end{equation}
where D is the distance in Mpc and K a correction factor for the
weighting of the source distribution by the antenna beam. When the
source was larger we used a factor 1.38 which leads to a main beam
scale. An exponential law was used for modelling radial distribution of
molecular gas with a scale length h=D$_B$/10, the molecular gas
following approximately the optical light distribution (Young \&
Scoville, 1982) i.e. the assumed gas surface density $\mu(r)$ is

\begin{equation}
\mu (r)\propto e^{-\frac{r}{h}}
\end{equation}

\begin{figure}
\centering
\psfig{width=8cm,height=6cm,file=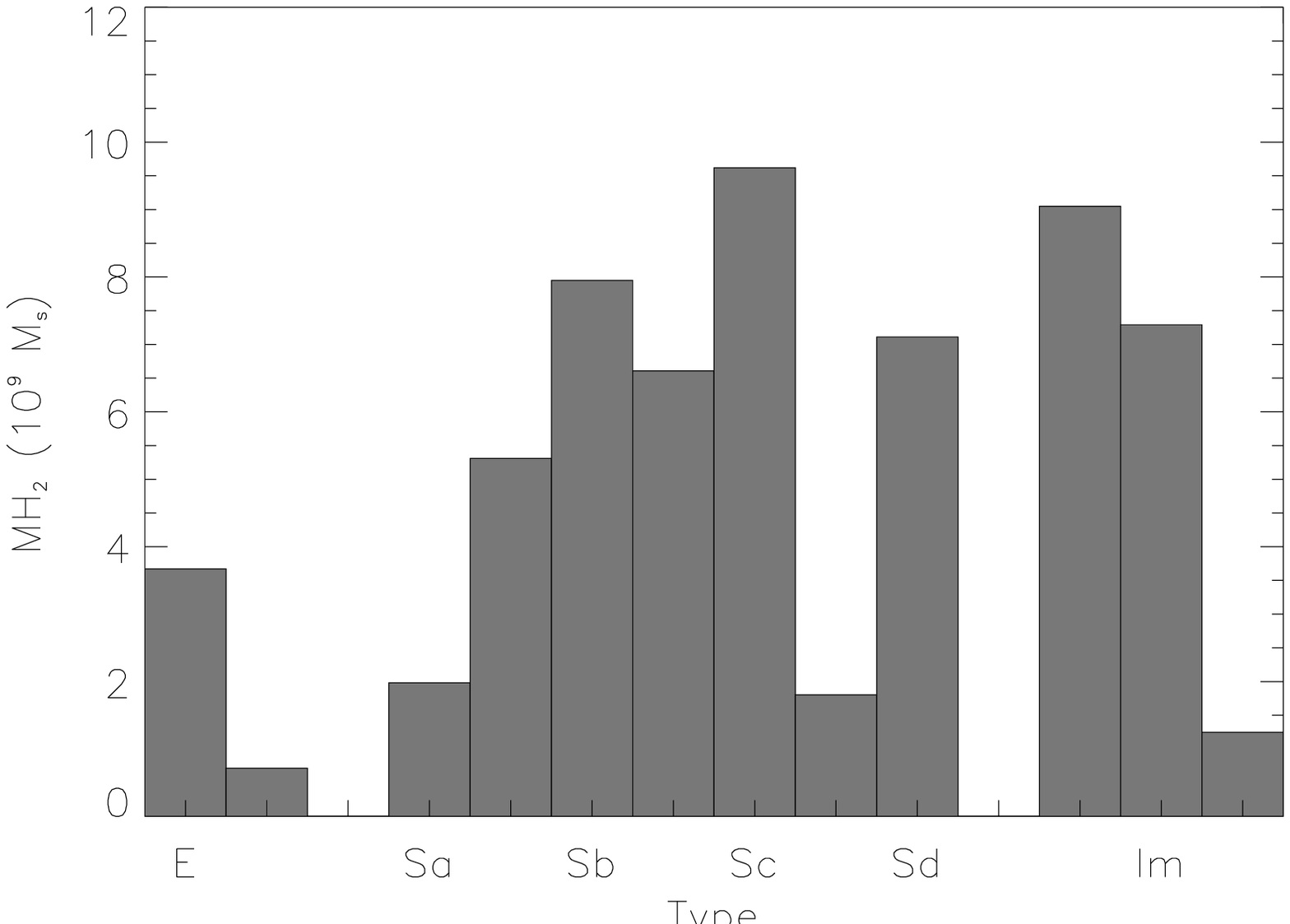}
 \caption{Mean \htwo\ mass ($10^9$ \msol) versus galaxy morphological type 
for our HCG sample.}
\label{gas_type}
\end{figure}

We could expect a more radially concentrated molecular distribution in
these tidally perturbed galaxies, but if we compare with a gaussian
distribution, there is at most a difference of a factor 1.5 on the
factor K. Thus M(H$_2$) will only be slightly overestimated through
this effect. We note  that this assumption works well for the group 16
where we have mapped in CO the whole galaxies.  For 75e and 80c a mean
ratio I$_{CO}$/I$_{CO(2-1)}$=1.35 is used to derive I$_{CO}$ intensity
in the \CO{1}{0} line. The mean \htwo\ mass versus galaxy morphological
type is presented in Fig. \ref{gas_type}.

\begin{table*}

\begin{tabular}{cccccccccc}
\hline
\hline                             
Name &  I$_{CO}$ & $\delta$I$_{CO}$ & $T_p$ & v$_{CO}$ & FWHM  & I$_{CO(2-1)}$  
 &Log(M(H$_2$))&Log(\lfir/\mhtwo)&Log(\lfir/($M_{gas}$))\\
   & (\Kkms)  & (\Kkms)  & (mK) & (\kms)         & (\kms)        & (\Kkms)  
   & (\msol)  & (\lo/\msol) &(\lo/\msol) \\
\hline	
2b  & 	3.90	& 0.25	& 37.8  & 4391	& 105		& 1.63		& 9.14	& 1.02&\\
3c  &	$<$1.80	& 	&  	&	&		& $<$2.02	& $<$9.17& &\\
4a  & 16.6	& 0.53	& 170.7	& 8047	& 83		& 		& 10.58& 0.38&\\
7a  & 15.70	& 0.80	& 54.5	& 4221	& 317		& 		& 10.02& 0.25&\\
7c  & 4.73	& 0.58	& 34.7	& 4401	& 124		& 1.51		& 9.59& 0.10& -0.23\\
10a & 2.72	& 0.49	& 9.3	& 5277	& 339		& $<$8.10	& 9.65& -0.63 & -1.08\\
10c & 7.14	& 0.36	& 21.8	& 4612	& 359		& 4.81		& 9.70& 0.14 & -0.12\\
11a$^{\dag}$& 1.84	& 0.23	&11.8 & 5433	& 171		& $<$2.70	& 9.41	& 0.23&\\
14b & 6.35	& 0.60	& 22.9	& 5867	& 383	& $<$2.50	& 9.60& &\\
14c & $<$1.50	&	& 	& 	&	& $<$3.60	& $<$8.76& &\\
16a & 42.86	& 1.29	& 	& 4056	& 		& 		& 9.98& 0.55 & 0.43\\
16b & 3.44	& 0.67	& 	& 3871	&		&		& 9.21&  &\\
16c & 62.95	& 1.53	& 	& 3835	&		&		& 10.15& 0.54 &\\
16d & 32.96	& 1.10	& 	& 3878	&		&		& 9.87& 0.88&\\
19b$^{\dag}$ & 1.23	& 0.30	& 8.1	& 4255		& 146		&		& 8.51& 0.86 &\\
21a & 14.86	& 1.20	& 53.0	& 7587	& 362		& 		& 10.31& 0.05 &\\
23b & 9.70	& 0.60	& 46.1	& 4914	& 318		&		& 9.85& 0.23&\\
23d & 2.15	& 0.20	& 24.8	& 4455	& 83		&		& 8.92& 0.37 &\\
25a & 2.56	& 0.20	& 13.3  & 6272	& 193		& 1.68		& 9.39& 0.55 &\\
25c & 6.03	& 0.49	& 21.8 	& 10894	& 394		& $<$1.62	& 10.07& 0.37 &\\
27b & 2.08	& 0.40	& 8.8	& 18516	& 311		&		& 10.01& 0.52 &\\
31a & $<$2.64	& 	&	&	&		&		& $<$8.84& &\\
31c & 1.53	& 0.30	& 15.7	& 3987	& 148		&		& 8.64& &\\
33c & 4.58	& 0.24	& 15.2	& 7787	& 362		& 19.1		& 9.63& 0.23&\\
34b & 5.99	& 0.70	& 19.7	& 9393	& 364		& 3.68		& 9.86& &\\
37b & 8.65	& 0.40	& 19.6	& 6763	& 552		& 5.93		& 10.00& -0.15 &\\
38b & 7.87	& 1.40	& 33.5	& 8677	& 251		& 		& 10.08&  &\\
40a & $<$1.32	&	&	&	&		& $<$2.88	& $<$8.90& &\\	
40c & 9.7	& 0.65	& 34.9	& 6375	&  233		&		& 9.94& 0.14 &\\
40d & 5.09	& 0.43	& 17.4	& 6729	&  295		& 3.09		& 9.57& -0.01 &\\
40e & 1.62	& 0.40	& 9.4	& 6480	&  66		& 1.23		& 8.98& 0.65 &\\
43a & 4.73	& 0.38	& 17.3	& 10026	& 256		& 1.49		& 9.86& 0.33 &\\
43b & 2.63	& 0.35	& 13.3	& 9920	& 278		& 1.82		& 9.66& 0.45 &\\
44a & 10.5	& 0.60	& 35.2	& 1239	& 269		& 3.65		& 9.13& -0.18& 0.06\\
44c & 7.63	& 0.50	& 50.2	& 1222	& 165		& $<$6.24	& 8.71& 0.23& 0.06\\
\hline
\end{tabular}
\caption{Molecular data}
\label{tab_co_1}
{$^{\dag}$not included in our final HCG sample}
\end{table*}

\addtocounter{table}{-1}
\begin{table*}

\begin{tabular}{cccccccccc}
\hline
\hline
Name &  I$_{CO}$ & $\delta$I$_{CO}$ & $T_p$ & v$_{CO}$ & FWHM  & I$_{CO(2-1)}$  
&Log(M(H$_2$))&Log(\lfir/\mhtwo)&Log(\lfir/($M_{gas}$))\\
   & (\Kkms)  & (\Kkms)  & (mK) & (\kms)         & (\kms)        & (\Kkms)  
   & (\msol)  & (\lo/\msol) &(\lo/\msol) \\
\hline	
44d & 1.79	& 0.19	& 18.5	& 1584	& 103		& 1.17		& 8.09& 0.55& -0.29\\
47a & 4.82	& 0.32	& 18.7	& 9630	& 321		& 1.23		& 9.92& 0.35 &\\
49a & $<$2.0	&	&	&	&		& $<$3.06	& $<$9.46& &\\
49b & $<$1.7	&	&	&	&		& $<$2.22	& $<$9.33& &\\
55a & $<$1.3	&	&	&	&		& 		& $<$9.68& &\\
57d & 1.45	& 0.24	& 10.0	& 9014	& 110		& 0.66		& 9.29& &\\
58a & 26.56	& 1.80	& 78.1	& 6132	& 510		&		& 10.41& 0.18& 0.10\\
59a & 6.05	& 0.25	& 44.2	& 4130	& 125		&		& 9.20& 0.96 & 0.63\\
59d & $<$1.40	& 	&	&	&		& $<$1.80	& $<$8.50& &\\
61c & 15.5	& 1.40	& 40.3	& 4034	& 484	& 7.50		& 9.82& 0.59 &\\
61d & $<$1.5	& 	& 	&	& 		& $<$2.00	& $<$8.51& &\\
67b & 6.69	& 0.46	& 19.9	& 7599	& 298		& $<$2.94	& 10.07& 0.21& -0.16\\
67c & 0.85	& 0.20	& 11.9	& 7485	& 64		& $<$1.05	& 8.87& &\\
68c & 7.02	& 0.42	& 32.2	& 2302	& 214		& 		& 9.49& 0.30& -0.09\\
69a & 4.3	& 0.55	& 12.5	& 8848	& 386		& $<$5.22	& 9.92& &\\
69b & 3.27	& 0.17	& 14.7	& 8728	& 279	& 4.47		& 9.58& 1.09&\\
71b & 4.88	& 0.65	& 24.3	& 9391	& 323		&		& 9.79& 0.82&\\
73a$^{\dag}$ & 2.85	& 0.45	& 28.7	& 5695	& 127		& 		& 9.44& -0.12& -0.87\\
75b & $<$2.40	& 	&	&	&		& $<$4.86	& $<$9.66& &\\
75e & $<$1.60	&	& 	& 12368	& 		& 0.53		& 9.14& &\\
78a$^{\dag}$ & 5.25	& 0.40	& 15.7	& 8628	& 363		& 5.00		& 9.97& 0.39& -0.10\\
79a & $<$1.70	& 	& 	&	&		& $<$2.52	& $<$8.68& &\\
79c & $<$2.10	&	&	&	&		& $<$3.90	& $<$8.77& &\\
80a & 10.40	& 0.60	& 29.7	& 9046	& 428		& 10.34		& 10.20& 0.64&\\
82c & 5.14	& 0.60	& 23.0	& 10090	& 388	& 2.96		& 10.03& 0.56&\\
88a & 5.96	& 0.80	& 16.2	& 5996  & 464	& 1.12		& 9.79& 0.22&\\
89c & $<$2.22	& 	&	& 8992	& 		& 0.56		& 8.96& &\\
92c & 4.95	& 0.70	& 17.2	& 6760	& 168		& 		& 9.97& 0.18& -0.09\\
93b & 9.95	& 0.52	& 46.5	& 4701	& 212		& 9.82		& 9.94& 0.31& 0.15\\
95b & $<$1.70	& 	&	&	&		& $<$2.28	& $<$9.55& &\\
95c & 3.96	& 0.30	& 17.7	& 11579	& 302	& $<$2.46	& 10.02& &\\
95d & 1.07	& 0.20	& 8.9	& 12074	& 173		& $<$1.80	& 9.34& &\\
96a & 17.4	& 1.10	& 116.5 & 8666	& 192 	& 13.74		& 10.57& 0.54&\\
96c & 2.69	& 0.40	& 11.4	& 8808	& 320	& 3.34		& 9.47& &\\
100a & 5.38	& 0.80	& 21.7	& 5298	& 402	& 		& 9.55& 0.72 & 0.31 \\
&\\
\hline
\end{tabular}
\caption{Molecular data (following)}
\label{tab_co_2}
{$^{\dag}$not included in our final HCG sample}
\end{table*}

\subsection{Average gas content and star formation efficiency}

 From Table \ref{tab_co_1}, average quantities can be derived to
characterize the HCGs as a class.
To avoid size effects and artificial correlations induced by uncertain
distances, we used quantities normalised to the blue luminosity \lb.
All the quantities are displayed in logarithm in the Table
\ref{average_Table}.  The average \lfir/\lb\ ratio is  -0.16 $\pm$ 0.45
for HCG, which indicates a moderate star forming enhancement over
isolated galaxies, as already found by Sulentic \& de Mello Raba\c{c}a
(1993). It should be emphasized here that the comparison among samples
using the blue luminosity normalised values of the total \lfir\ and
\mhtwo\  by Sulentic \& De Mello Raba\c{c}a (1993) assumes that the
spatial distributions of these two quantities are similar to the blue
luminosity distributions. However as stressed by Menon (1995) in his
comparison of radio properties of HCG spirals and isolated spirals the
differences in spatial distributions have to be taken into account for
any meaningful comparison of sample properties.  The average
\mhtwo/\lb\ ratio is found to be -0.61 $\pm$ 0.39, which has to be
compared with the -1.19 value of Boselli et al.  (1996). This
difference might be due to the small size of the Boselli et al. sample and 
also to the fact that
they do not take into account the correction of galaxy-to-beam size
ratio in deriving the \htwo\ content. Yun et al (1997) also reported an
apparent CO emission deficiency in two HCG groups that they mapped with
the OVRO interferometer (31c and 92c).  However, their observations are
missing extended CO emission, and they find 2 and 10 times less
\mhtwo\ than the present work, for 31c and 92c respectively.

\begin{figure}
\centering
\psfig{width=8cm,height=6cm,file=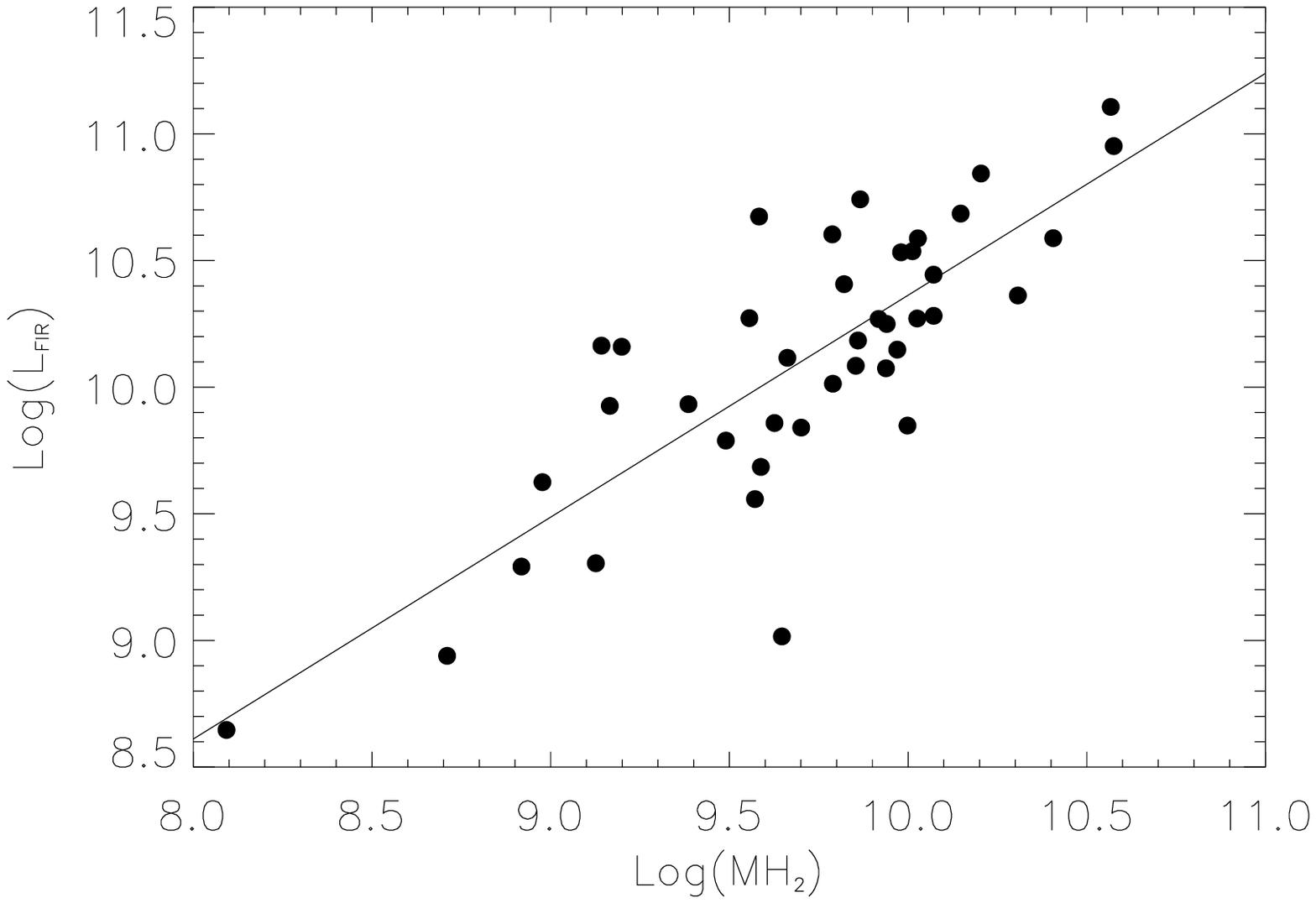}
\caption{The FIR luminosity versus the \htwo\ mass for our HCG sample. The 
least square linear fit is superposed.}
\label{mh2_fir_correl}
\end{figure}

The \lfir/\mhtwo\ ratio is widely used as an indicator of the star
formation efficiency (SFE) (Young et al. 1986). While our present
control has an average SFE of 0.67 $\pm$ 0.38, the HCG sample has an
SFE of 0.39 $\pm$ 0.33, which confirms the only moderate triggering
effect of the compact environment on the global star formation.This 
value has to be compared with the high ratio of 1.24 for the starburst 
sample which is about 7 times higher. If we 
take as indicator of SFE the ratio \lfir/(\mhtwo+\mhi), as proposed by
Sulentic \& de Mello Raba\c{c}a (1993), no star formation enhancement
is observed with a mean ratio of -0.02 $\pm$ 0.40; however, this value
is then uncertain, due to the available sample of only 14 galaxies with
HI content known, due to the poor spatial resolution of the
\hi\ observations (Williams \& Rood, 1987) and the lack of interferometric
observations up to now. As already observed, we find a close correlation
between the absolute \mhtwo\ content and the \lfir.  This correlation
is usually interpreted as a relation between the fuel for star
formation (molecular gas) and the tracer of that star formation (FIR
luminosity) (Young et al., 1986).  On Fig. \ref{mh2_fir_correl} we
present \lfir\ versus \mhtwo\ superposed with the linear fit (in log):
\lfir=40\mhtwo$^{0.88\pm 0.31}$. One of the galaxies (10a) exhibits a
particular high \htwo\ content, without any counterpart of high \lfir\
luminosity, that could be due to a peculiar (not exponential-like) gas
distribution, which would imply that our estimation of the total gas
fails.  Indeed that galaxy has a very large diameter relative to our
beam. Mendes de Oliveira \& Hickson (1994) remarked that it exhibits a
peculiar \hi\ profile.

\begin{table*}

\begin{flushleft}
\caption{Average quantities$^a$ with their standard deviation.}
\label{average_Table}
\scriptsize
\begin{tabular}{rrrrrrrr}
\hline
\hline
 & & & & & & \\
\multicolumn{1}{c}{Sample}                                    &
\multicolumn{1}{c}{\mhtwo/\lb}                                 &
\multicolumn{1}{c}{$L_{\rm FIR}/L_{\rm B}$}                    &
\multicolumn{1}{c}{$L_{\rm FIR}/M_{{\rm H}_2}$}                &
\multicolumn{1}{c}{$L_{\rm FIR}/(M_{{\rm H}_2}+$\mhi)}         &
\multicolumn{1}{c}{(\mhtwo+\mhi)/\lb}                         &
\multicolumn{1}{c}{$M_{\rm d}$/\lb}                            &
\multicolumn{1}{c}{$T_{\rm d}$}                                \\
 & 
\multicolumn{1}{c}{\hbox{${\rm M}_{\odot}/{\rm L}_{\odot}$}}   &
&
\multicolumn{1}{c}{\hbox{${\rm L}_{\odot}/{\rm M}_{\odot}$}}   &
\multicolumn{1}{c}{\hbox{${\rm M}_{\odot}/{\rm L}_{\odot}$}}   &
\multicolumn{1}{c}{\hbox{${\rm M}_{\odot}/{\rm L}_{\odot}$}}   &
\multicolumn{1}{c}{\hbox{K}}                                \\
 & & & & & & \\
\hline
 & & & & & & \\
HCG 	&  -0.61(0.39) & -0.16(0.45) & 0.39(0.33) & -0.02(0.40)  & -0.42(0.22) & -3.42(0.36) &  33.1(5.7)  \\
 & & & & & & &\\
Control &  -0.78(0.58) & -0.11(0.61) & 0.67(0.38) & 0.06(0.41)   & -0.34(0.35) & -3.54(0.41) &  35.0(5.4)  \\
& & & & & & & \\
Pairs   &  -0.57(0.45) &  0.33(0.48) & 0.91(0.43) & 1.04(0.37)   & -0.47(0.31) & -3.37(0.40) &  34.9(6.0)  \\
& & & & & & &\\
Starburst & -0.61(0.43)&  0.63(0.43) & 1.24(0.39) & 0.91(0.39)	 & -0.36(0.40) & -3.27(0.34) &  40.4(6.2)  \\
& & & & & & &\\
Cluster	&  -1.08(0.36) & -0.31(0.40) & 0.77(0.37) & 0.42(0.35)	 & -0.73(0.34) & -3.78(0.30) &  33.2(4.7)  \\
& & & & & & &\\
Dwarf	&  -1.65(0.88) & -0.34(0.59) & 1.38(0.68) & 0.13(0.43)   & -0.32(0.65) & -4.11(0.67) &  38.1(7.2)  \\
& & & & & & & \\
Elliptic&  -1.65(0.77) & -0.46(0.62) & 1.19(0.48) & 0.39(0.32)   & -0.64(0.24) & -4.07(0.63) &  33.1(5.2)  \\
\hline
\end{tabular}

a)\ All averages are logarithmic

\end{flushleft}

\end{table*}


\subsection{Dust masses}

\begin{figure}
\centering
\psfig{width=8cm,height=6cm,file=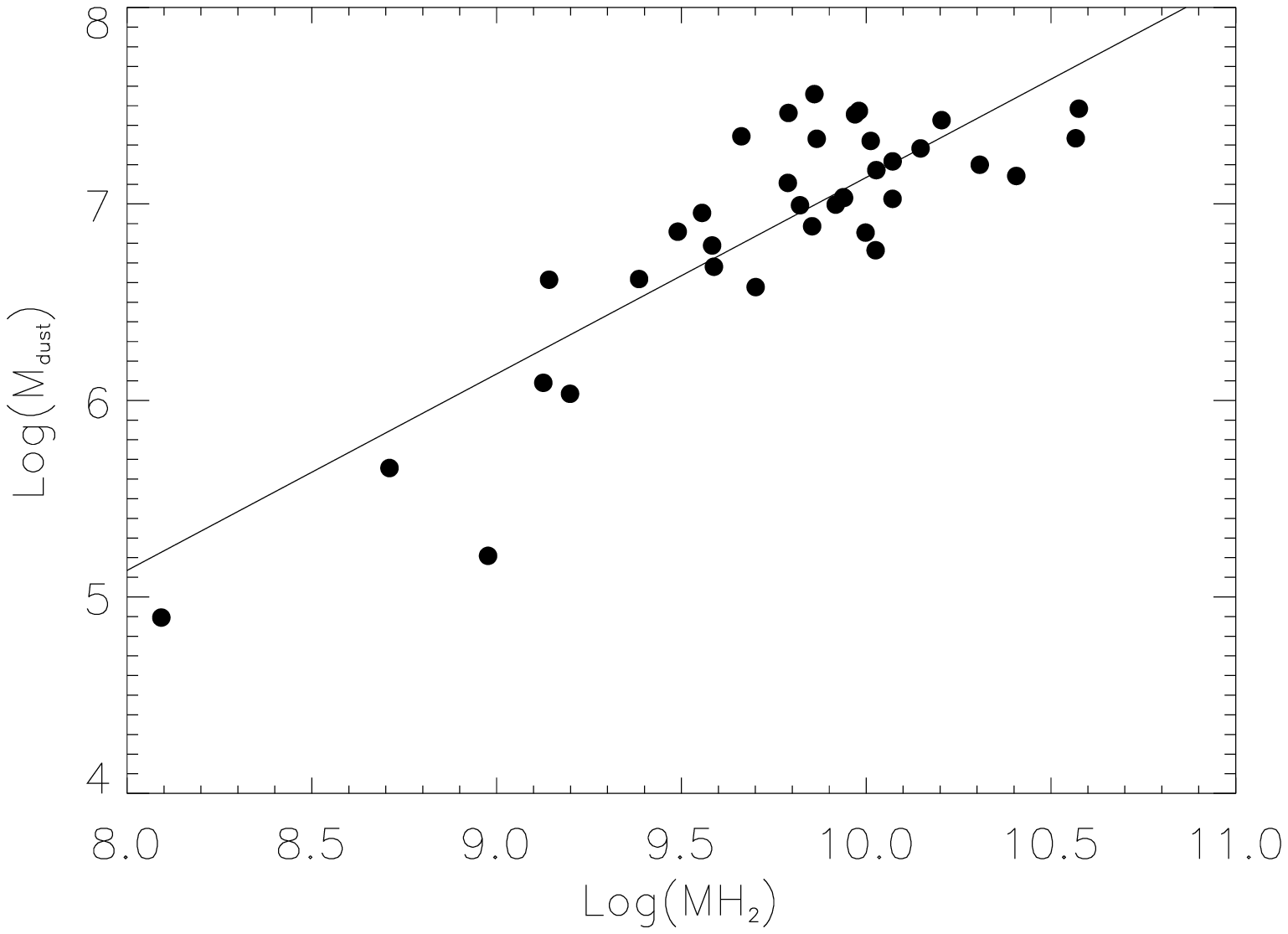}
\caption{The molecular mass derived from the CO intensities for
our HCG sample, versus the dust mass derived from the FIR luminosity.
The full drawn line represents a simple proportionality between
the two mass estimates for a molecular gas--to--dust mass ratio of 725.
}
\label{dust_h2}
\end{figure}

From the ratio of the IRAS \ssix\ and \shun\ fluxes we have derived
dust temperatures (cf Table \ref{tab_sample_1}),  assuming
$\kappa_{\nu} \propto \nu$.  The average \tdust\ for the HCG galaxies
with detected CO emission is 33\,$\pm$\,6\,K.  By comparison, the
average dust temperature for the starburst sample is 40\,$\pm$\,6\,K.

Knowing the dust temperature \tdust\ and the \shun\ flux,
we can derive the dust mass as
\begin{eqnarray*}
  M_{\rm dust} & = & 4.8 \times 10^{-11}  \,
  {S_{\rm \nu}\,d_{\rm Mpc}^{\,2}
  \over \kappa _{\rm \nu }\,B_{\rm \nu}(T_{\rm d})}\ {\rm M}_{\odot } \\
  & = & 5 \,S_{100}\,d_{\rm Mpc}^{\,2}\,
  \left\{\exp (144/\tdust) - 1 \right\}\ {\rm M}_{\odot },
\end{eqnarray*}

\begin{figure*}
\centering
\psfig{width=14cm,height=10cm,file=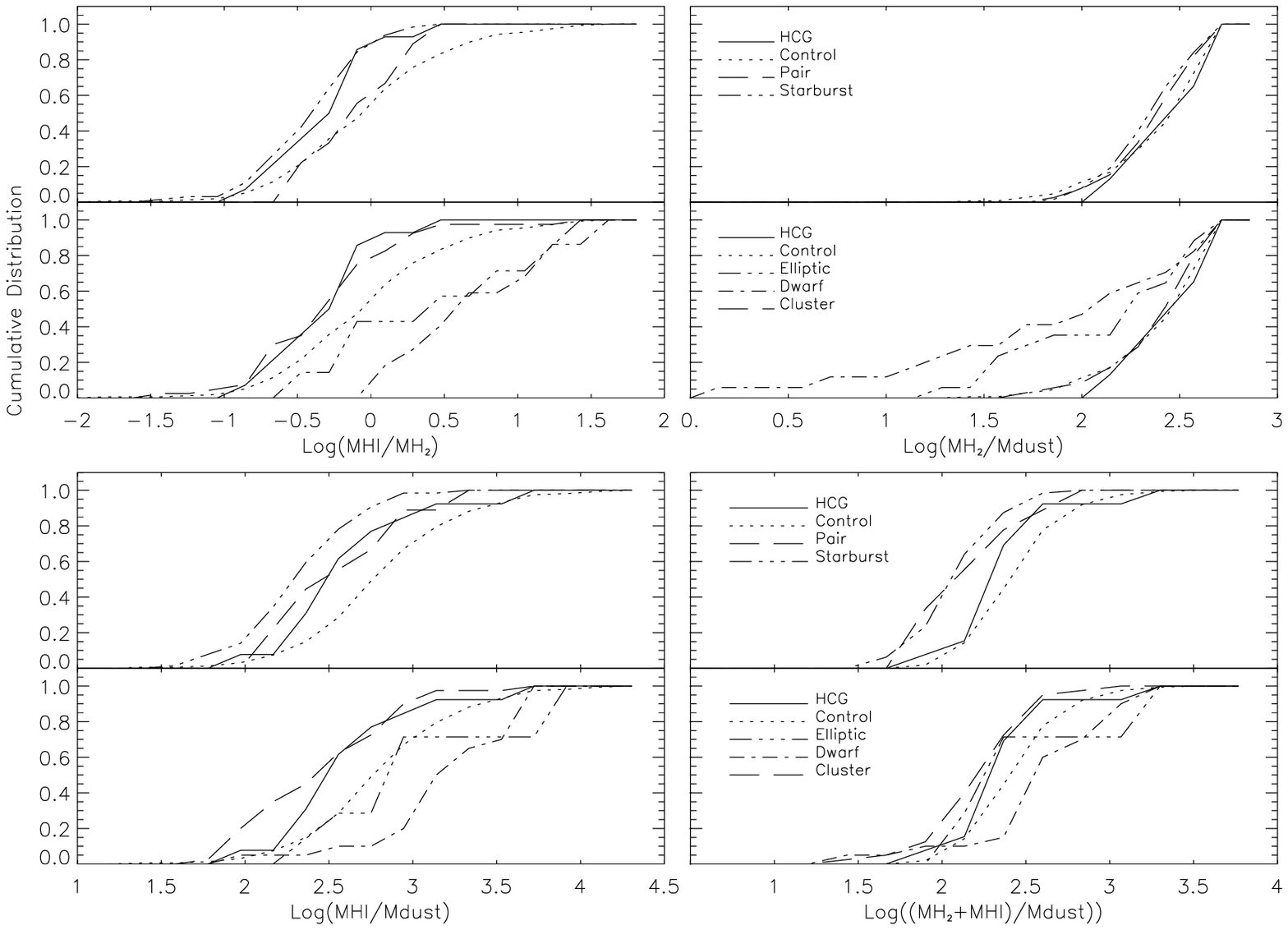}
\caption{The cumulative distributions for the different samples described in the
text. Here are represented the ratio of the \htwo, \hI\ and dust masses.}
\label{df_ratio}
\end{figure*}

where $S_{\rm \nu}$ is the FIR flux measured in Jy, $\kappa _{\rm \nu}$
is the mass opacity of the dust, and $B_{\rm \nu}$(\tdust)\ the Planck
function. We used a mass opacity coefficient of 25\,cm$^{2}$\,g$^{-1}$
at 100$\mu$ (Hildebrand, 1983).  In Table \ref{tab_sample_1} we list
the estimated dust masses, and in Fig.  \ref{dust_h2} we plot the dust
mass derived from the FIR flux versus the \htwo\ mass derived from the
CO observations. The full line is a fit of the dust mass to the
\htwo\ mass, corresponding to a simple proportionality between the two
masses, with a molecular gas--to--dust mass ratio of 725 which is similar 
to the molecular gas--to--dust ratio of the control sample (741). 
The total gas--to--dust
ratio is not reliable (high dispersion) given our small  sample with
FIR and \hi\ available.  
We computed for all our samples the cumulative distribution of
the ratio \mhtwo/\mdust (log). It is clear from Fig. \ref{df_ratio} that
this ratio in all samples, except the dwarf and elliptic samples ,
follows the same distribution, the maximum \chitwo\, 
with the HCG sample being 4.1
(probability 0.46) to be compared with the Table \ref{table_test}. 
Although there is some observational
evidence in favor of a constant gas-to-dust ratio in galactic giant
molecular clouds (GMCs), as shown by Sanders et al. (1991), a departure
from this value could indicate that either a fraction of the FIR
luminosity comes from dust associated with the diffuse atomic ISM or
that the \htwo/CO conversion factor is galaxy-dependent.  Sanders et
al. (1991) also suggest that the opacity coefficients could
underestimate dust masses.

\subsection{The CO(2-1)/CO(1-0) ratio}

 We detected the \CO{2}{1} line in 26 galaxies only (55\% detection
rate). In the other CO detected galaxies, we have only an upper limit
of the CO(2-1)/CO(1-0) ratio.  In most cases, we observed only one
position per galaxy: since the beam sizes are different for the two
lines, we cannot determine the true line ratio, without a precise model
of the source distribution. In any case, we measure an average raw line
ratio of 0.74 $\pm$ 0.2, without beam correction. The true ratio will
be obtained by dividing by a factor between 1 and 4, because of the
factor 2 between the CO(2-1) and CO(1-0)  linear beam size. It is thus
certain that the CO emission is in general sub-thermally excited, as is
frequently the case at large scale in galaxies (e.g. Braine \& Combes
1992). There is only one exception, the galaxy 33c, where the true
CO(2-1)/CO(1-0) ratio could be of the order of 1. It has been shown
that the CO line ratio varies little with the interaction class of the
galaxy (Casoli et al. 1988, Radford et al. 1991); it is not a good
temperature indicator, but rather a density indicator (it is higher in
the galaxy centers, as expected). As a matter of fact this global 
ratio cannot disentangle in any way the different excitation CO 
conditions in the galaxies (hot cores, diffuse component).

\subsection{Correlation with the radio continuum flux}

\begin{figure}
\centering
\psfig{width=8cm,height=6cm,file=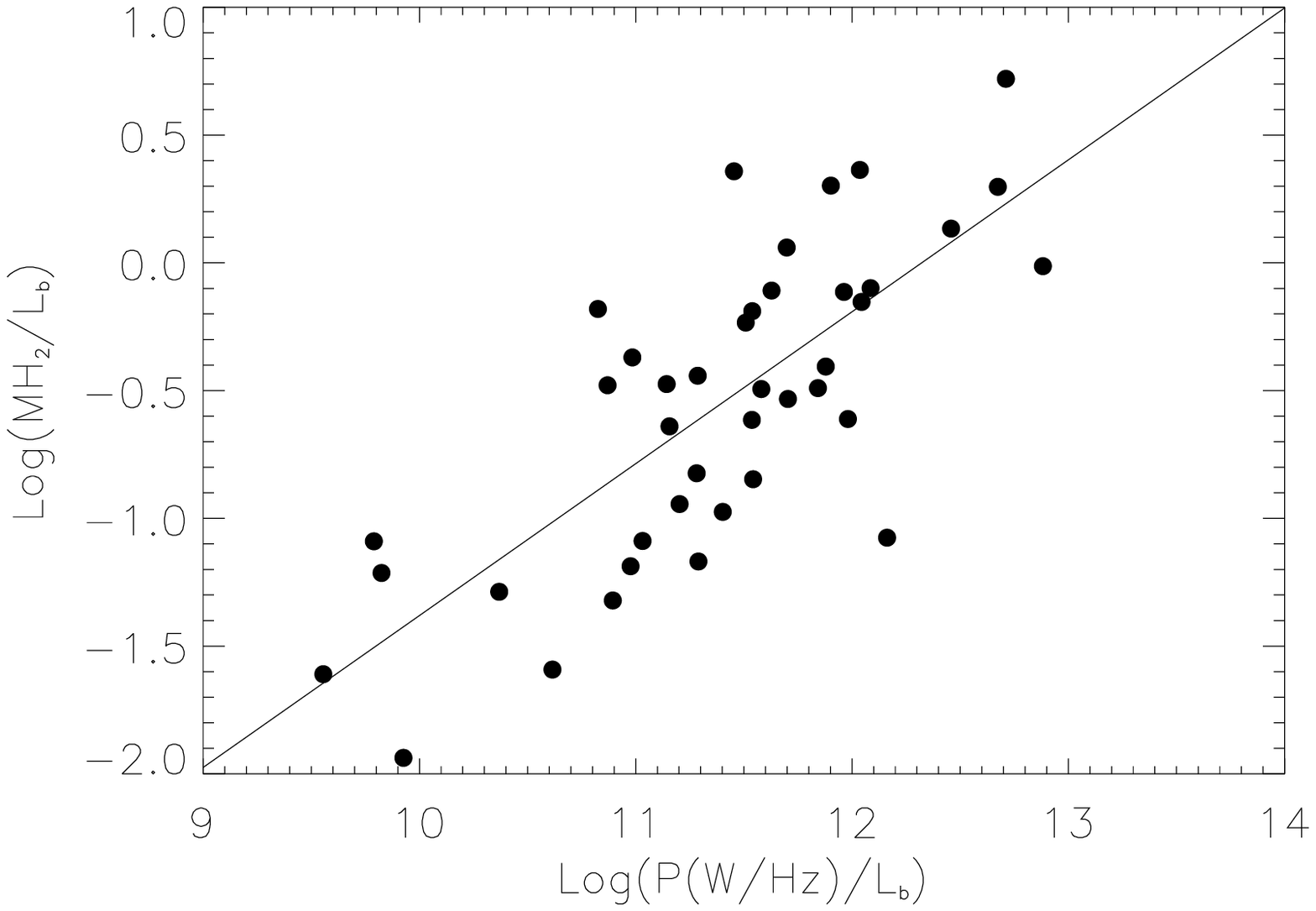}
\caption{The molecular mass, normalised to \lb,  of HCG galaxies versus their
radio continuum emission. The least square linear fit is overplotted.}
\label{radio_CO}
\end{figure}

 We have plotted \mhtwo\ versus the  radio luminosity at 1.4 GHz  of 
the detected compact
group galaxies in Fig. \ref{radio_CO}. A clear correlation can be 
found, indicating that both are related to star-forming activity.
Menon (1995) has shown that the total radio emission from the discs of
HCG galaxies is significantly less than that of a comparable sample of
isolated galaxies, while the reverse is true for the nuclear emission.
He suggested that the nuclear radio emission is mainly due to star
formation bursts and not due to nuclear activity. This is supported
here from the good correlation between normalised molecular gas content
and radio power. 
AGN-powered radio emission should perturb this
correlation, but this appears negligible here.

\subsection{CO maps of a few objects}

\begin{figure*}
{\centering
\psfig{width=14cm,height=10cm,file=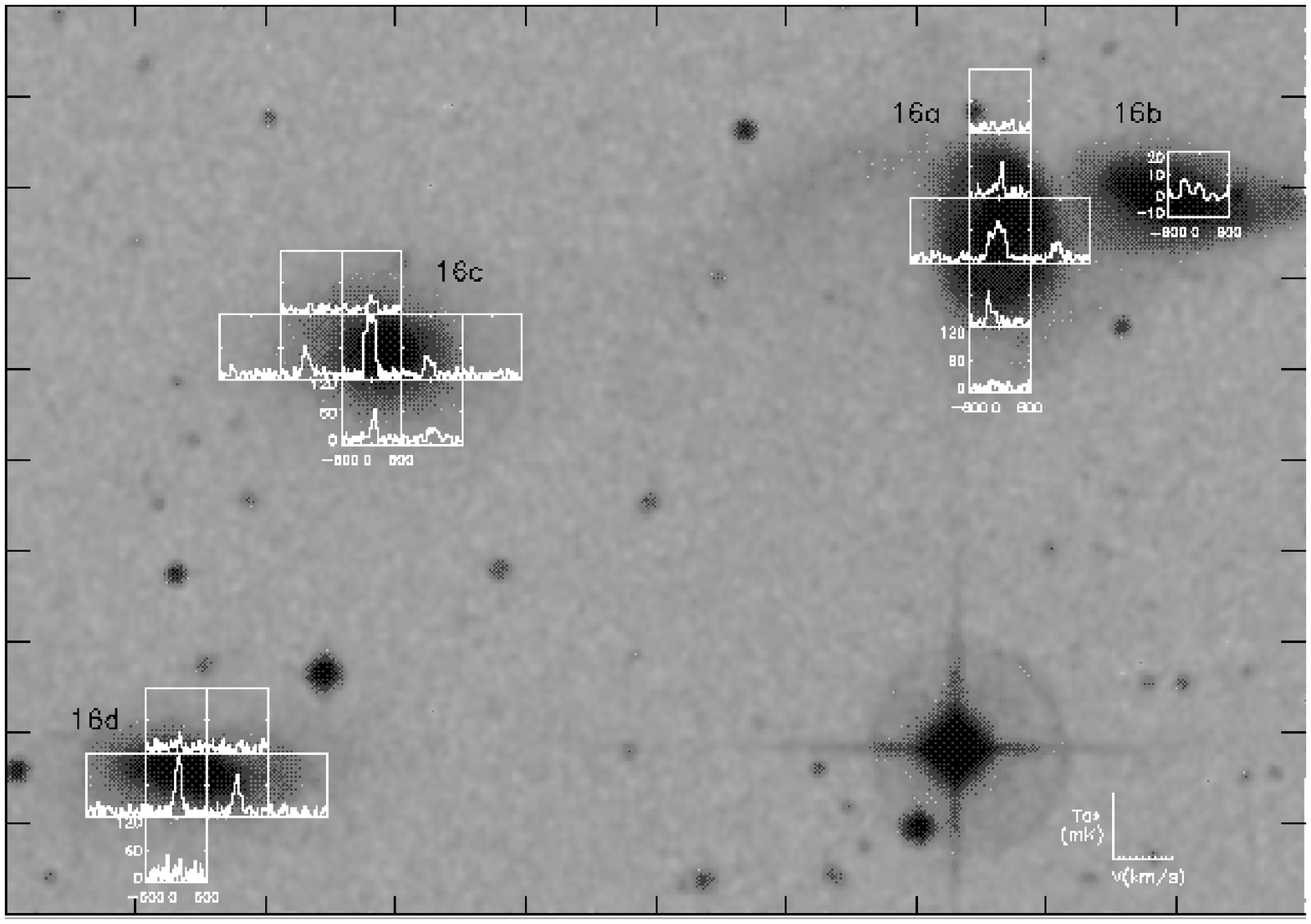}}
\caption{Spectra towards the HCG16 galaxies, superposed on 
an optical image of the group (taken from the DSS). The velocity is in \kms
centered on the recession velocity of each galaxy and the y-axis is in
units of \tastar (mK). }
\label{HCG16}
\end{figure*}

Some of the Hickson groups are near enough to be resolved by our beam,
and we mapped a few objects, in particular several galaxies in HCG16.
This compact group appears as a unique condensation of active galaxies,
containing one Seyfert 2 galaxy, two LINERS and three starbursts
(Ribeiro et al 1996). The galaxy density is 217 gal Mpc$^{-3}$. Ponman
et al (1996) detected a diffuse X-ray component corresponding to
intra-cluster hot gas.  We present in Fig. \ref{HCG16} the CO spectra
towards the HCG16 galaxies. Some of them (16a and 16b) reveal several
velocity components, which can be attributed to overlapping galaxies.
16c and 16d present a clear enhancement of their molecular content with
an \mhtwo/\lb\ ratio equal respectively to -0.21 and -0.37 (log). These two
galaxies exhibit optical starburst activity which indicates recent
interaction, younger than $10^8$ yr (Ribeiro et al.  1996); while the
galaxies 16a and 16b show tidal tails indicating a later interaction
phase.  They should already have suffered intense star formation and
have consumed part of the fueling gas available.  To our spatial
resolution of 22\arcsec\ we do not observe such a high central
concentration of the molecular gas as for radio continuum emission
(Menon, 1995).

\subsection{About the compactness}

\begin{figure}
\centering
\psfig{width=8cm,height=6cm,file=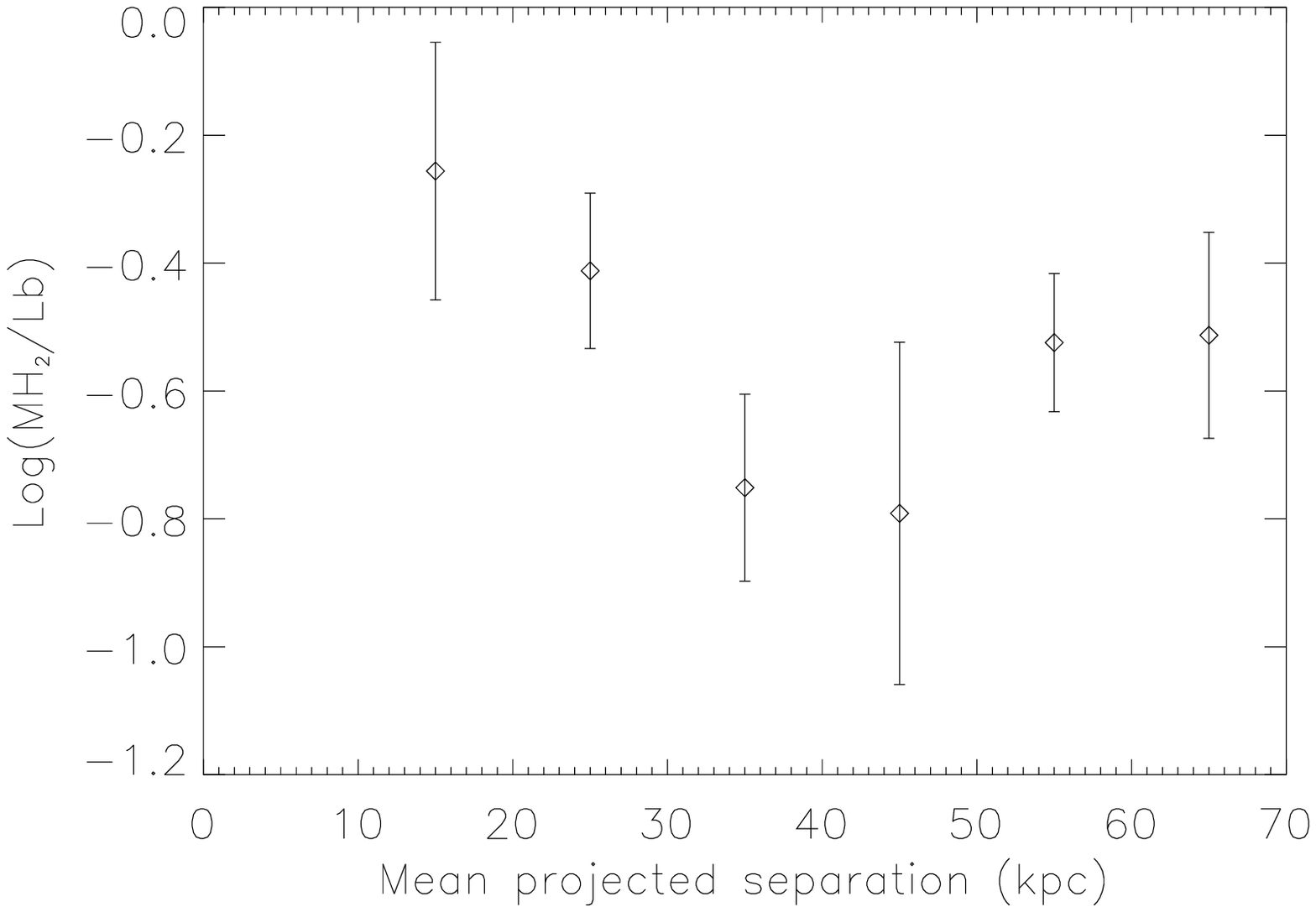}
\caption{\mhtwo/\lb\ ratio versus the mean separation in each group 
per bin of 10 kpc.}
\label{mh2_separation}
\end{figure}

We plot in Fig. \ref{mh2_separation} the mean \mhtwo/\lb\ ratio
versus the mean projected separation in the group. There is an
enhancement of the \htwo\ content up to a mean separation of 30 kpc.
This is an indication for the interaction intensity threshold to
trigger inner gas flows by tidal interactions.  Galaxies with a mean
separation less than 25 kpc have \mhtwo/\lb=-0.26 ($\pm$0.33), whereas
galaxies with separation more than that distance have \mhtwo/\lb=-0.66
($\pm$0.37). It is interesting to note that the correlation is weaker
when the closest projected separation is used instead of the mean
separation in the group.  This suggests that the enhancement is a
function of the dynamics of the {\em whole} group in the case of the
most compact groups, apart from possible strong binary interactions, as
shown in HCG16. The \lfir/\mhtwo\ ratio does not exhibit any dependence
with the mean separation.

\section{Discussion}

\subsection{Gas content}

\begin{figure*}
\centering
\psfig{width=14cm,height=10cm,file=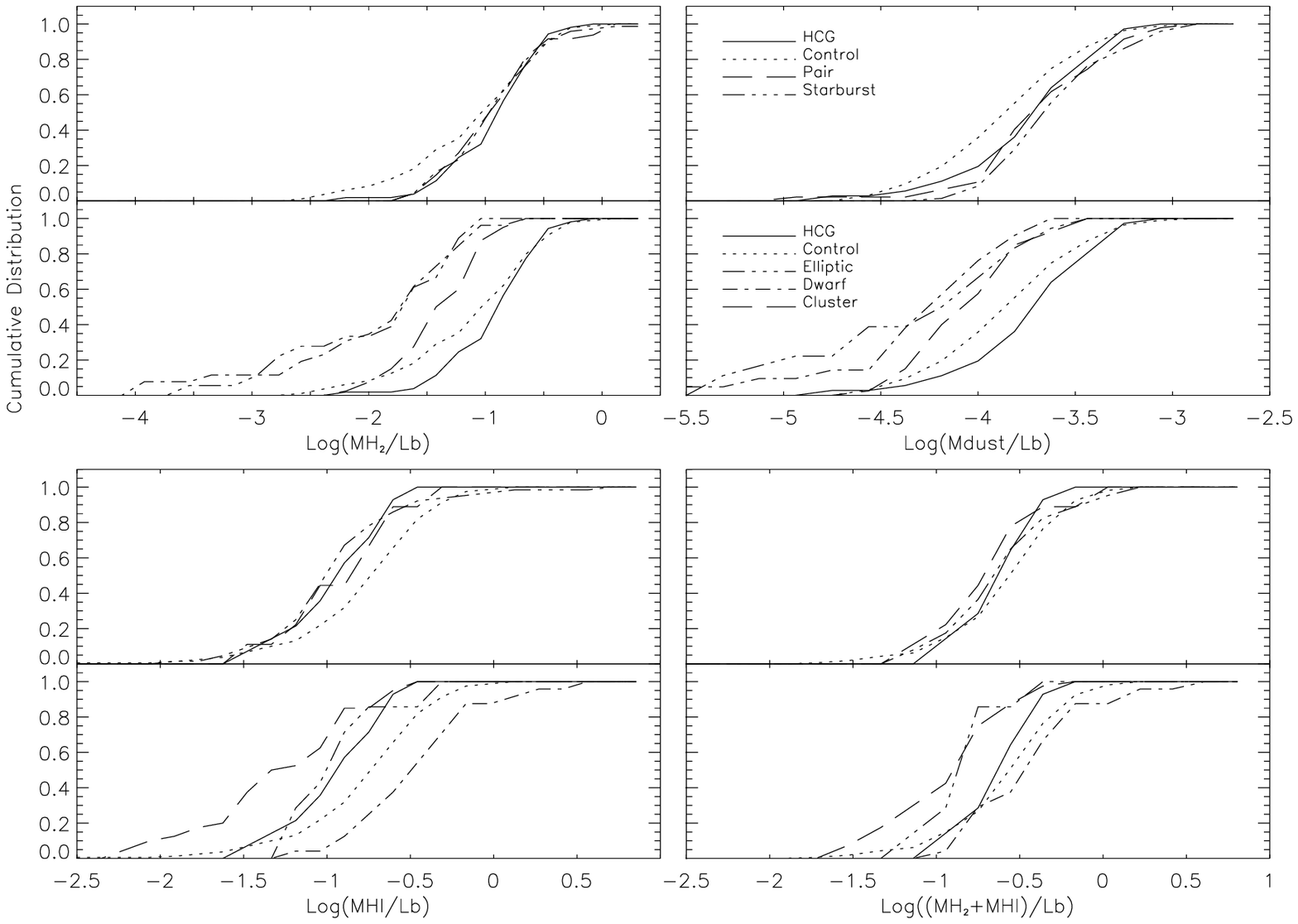}
\caption{Cumulative distribution of the different samples for the \htwo, \hI\
and dust masses. The top graphs compare the HCG with control, pair and
starburst ensembles; the bottom ones to the other categories. }
\label{df_mass}
\end{figure*}

\begin{figure*}
\centering
\psfig{width=18cm,height=14cm,file=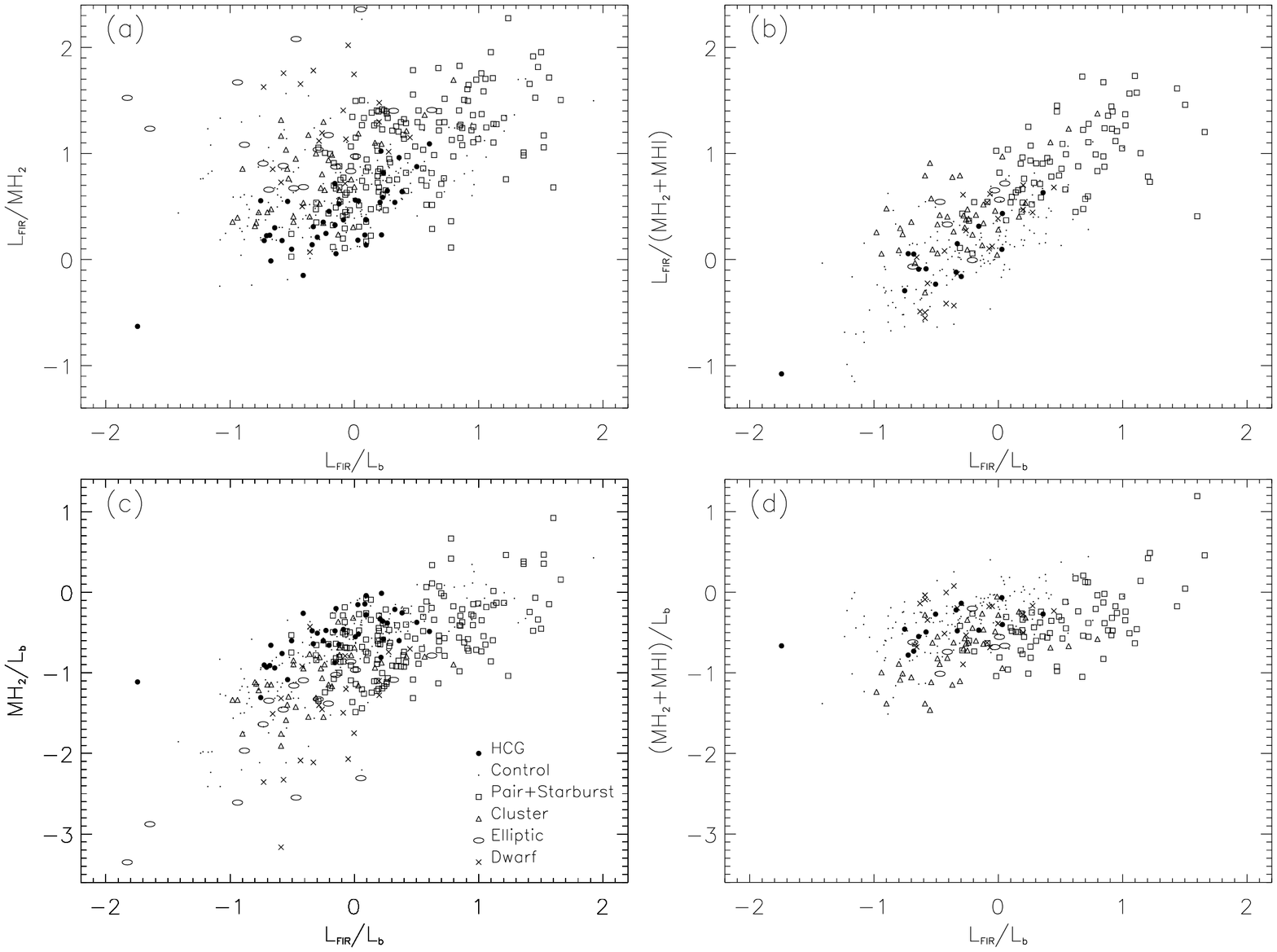,angle=90}
\caption{Comparison between our HCG sample and other CO-detected samples:
(a)-- Star Formation Efficiency (SFE) as represented by 
\lfir/\mhtwo\ versus \lfir/\lb.
(b)-- same as a) but including the HI gas in the definition of the SFE
(c)-- \mhtwo\ versus \lfir\ normalised to blue luminosity.
(d)-- total gas content (\mhtwo\ +M(HI)) versus \lfir\ normalised to blue 
 luminosity.
}
\label{fir-H2-SFE-HI}
\end{figure*}

To compare our HCG sample with the comparison samples we used the
cumulative distributions for the different quantities. Performing
Kolmogorov-Smirnov (KS) or \chitwo\ test we check the hypothesis of a
common underlying population for the different samples.

One of the result derived from the comparisons of Fig. \ref{df_mass} 
concerns the CO emission: there does seem to be an enhancement of \mhtwo\  
 in HCG galaxies with respect to a control sample which does not share the same underlying 
distribution (KS: 0.21 (0.05 significance), \chitwo: 15.66 (0.64 probability)). 
However HCG population seems to share the
same distribution with starburst (KS: 0.12 (0.61), \chitwo: 9.65 (0.68)) and pair 
(KS: 0.22 (0.16), \chitwo: 11.30 (0.58)) samples. We present in Table
\ref{table_test} the KS and \chitwo\, results for all coupled samples.
The distribution functions of the HCG and control sample exhibit the main 
difference for the low H$_2$ content galaxies. 
Although the
HCG sample does not exhibit a global FIR enhancement for all galaxies, as
shown in Fig. \ref{df_fir}, it appears that tidal interactions are 
efficient in Compact Groups, at least in the most compact ones as 
we have shown in the previous section. These tidal torques could drive the gas
inwards, which might be related to the
enhancement of radio continuum emission in the very center of these
galaxies. The \htwo\ enhancement does not appear to be  a bias from
our FIR selected sample since it has a FIR distribution close to that
of the control sample (KS: 0.14 (0.49)) but we will discuss
afterwards about the Malmquist bias which could be an important 
limitation in this issue. We can point out that in case of a perturbed
 molecular gas distribution, our extrapolation for the
total mass should fail. But in this case the conclusion would remain similar,
leading that time to an enhancement of the molecular content in the
{\em center}. This enhancement of molecular gas is also supported by
the enhancement of dust mass. Its cumulative distribution in HCG also
follows that of the starburst (KS: 0.18 (0.38),\chitwo: 8.65 (0.69)) and pair
(KS: 0.11 (0.91),\chitwo: 9.08 (0.72)) samples.  In spite of the poor \hI\ data, we can
point out that the similarity between these three populations (HCG,
pairs and starbursts) for the total gas (\mhtwo+\mhi)/\lb\ ratio is
even tighter (KS: 0.12 (0.99), \chitwo: 5.51 (0.79), gathering pair and starburst samples 
in one sample). The case of dwarfs,
exhibiting a very peculiar molecular gas-to-dust ratio, can be interpreted in
terms of the low metallicity of these objects and will be discussed in
a forthcoming paper (Leon et al., 1997).\\
Thus in a galaxy group two mechanisms are at play concerning the gas evolution:
on one hand tidal interactions enhance the molecular content by driving gas 
inwards and on the other hand the Intra Cluster Medium strips off the 
outer gas, reducing the eventual molecular content.
It appears that the former dominates for the most compact groups, while for the
least compact groups the picture is mitigated. Nevertheless it can be 
emphasized that the enhancement of dust mass gives us a hint about 
evidence of tidal interactions in the HCG sample.

\begin{table*}
\caption{KS and \chitwo\, statistics for the $\log(\mhtwo / \lb )$ 
distribution between every two different samples. In the upper part we give 
the  \chitwo\, value and the probability (brackets) 
of the test, in the lower part is the maximum difference between two 
cumulative distributions and the KS significance (brackets).}
\label{table_test}
\begin{tabular}{c|ccccccc}
 & 	  	HCG & 	control &  starburst & 	pair &  	cluster & 	dwarf &    elliptic \\
\hline
HCG &  		    & 15.66(0.64) & 9.65(0.68)    & 11.30(0.58) & 34.36(0.02)  & 47.75(0.01) & 40.61(0.02) \\
control & 0.21(0.05)&		  & 19.08(0.52)	  & 22.65(0.33) & 32.00(0.10)  & 57.55(0.00) & 63.06(0.00) \\
starburst &0.12(0.71)&0.16(0.14)  & 		  & 9.98(0.66)  & 32.49(0.04)  & 52.06(0.00) & 49.10(0.00) \\
pair  &	0.22(0.16)  & 0.18(0.15)  & 0.10(0.95)	  & 		& 25.11(0.08)  & 39.22(0.03) & 34.73(0.04) \\
cluster&0.59(0.00)  & 0.40(0.00)  & 0.51(0.00)	  & 0.50(0.00)  &	       & 24.88(0.19) & 18.96(0.26) \\
dwarf &	0.66(0.00)  & 0.50(0.00)  & 0.65(0.00)    & 0.67(0.00)	& 0.41(0.01)   &	     & 16.90(0.49) \\
elliptic & 0.68(0.00)& 0.53(0.00) & 0.66(0.00) 	  & 0.61(0.00)	& 0.33(0.11)   & 0.18(0.89)  &		   \\
\end{tabular}
\end{table*}

\subsection{Completeness of the distribution}

\begin{figure}
\centering
\psfig{width=8cm,height=6cm,file=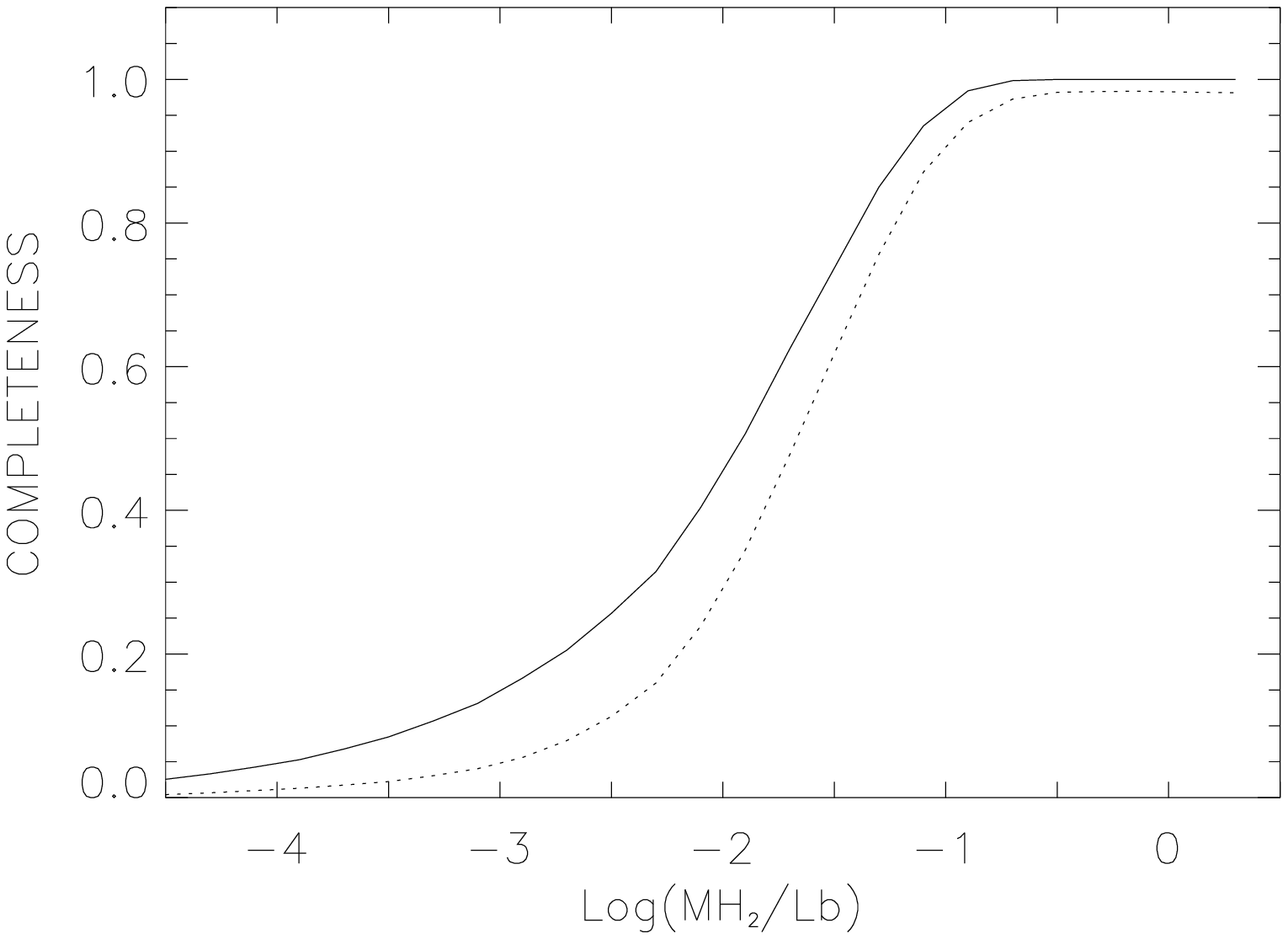}
\caption{Completeness distribution  of the $\log(\mbox{\mhtwo / \lb} )$ 
quantity for our sample. Uniform (solid line) and gaussian (dashed line) 
distribution for the simulated distance in our realisations
are shown.}
\label{plot_completude}
\end{figure}

We check the completeness of the \mhtwo/\lb\ distribution function 
(top left of Fig. \ref{df_mass}) by simulating the distribution
function taking into account the threshold of detection for that quantity. 
The minimum temperature detection is 4 mK in our observations.
Then we consider the distance distribution for the galaxies to be uniform
up to 150 Mpc or gaussian with the parameters of our sample. 
The linewidth distribution has been fitted to the blue luminosity \lb\
with a power law (Tully-Fisher-like relation with $\lb \propto {\Delta V}^{4.3}$). 
Then inclination angle is distributed uniformly between 0 and 90 degrees. 
The blue luminosity is distributed
 with a gaussian distribution ($<\log (\lb ) >=10.22$, $\sigma_{\lb}=0.16  $).
For each $\log ( \mhtwo / \lb ) $ bin, the fraction of realisations above the 
threshold detection is computed
to estimate the completeness of our sample. 
Results are displayed in Fig. \ref{plot_completude} for $10^4$ realisations.
Gaussian and uniform distance distributions are two extremes
chosen to estimate the weight of the
distance parameter: the 50 \%  level of completeness is 
$ \log ( \mhtwo / \lb ) =-1.8 $ for the uniform case and -1.5 for the gaussian case.

\begin{figure}
\centering
\psfig{width=8cm,height=6cm,file=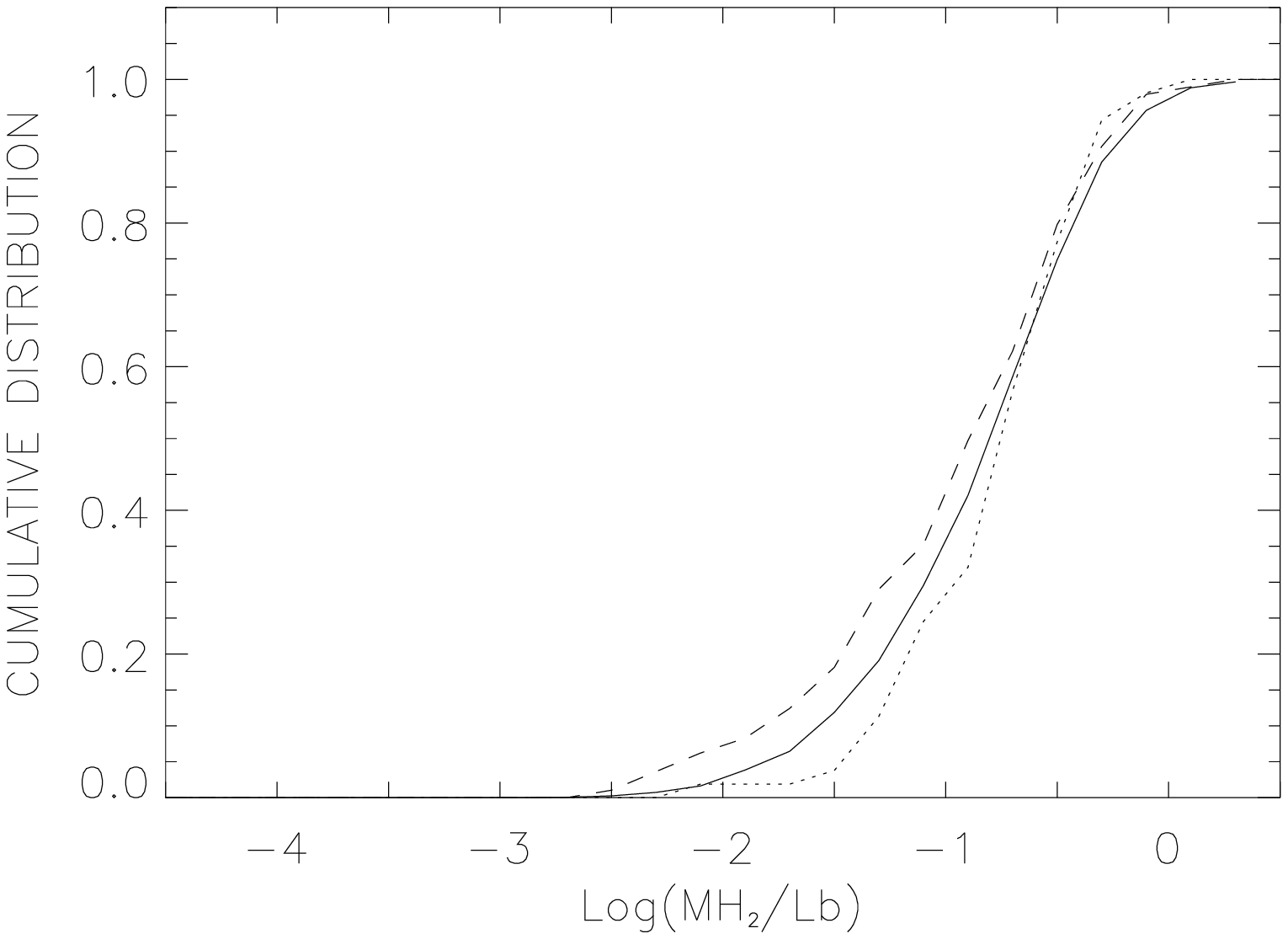}
\caption{Cumulative distribution (solid line) of the $\log(\mbox{\mhtwo 
/ \lb} )$ quantity simulated from the completeness of the distribution (see 
text) and the observed distribution
for the HCG (dot line) and the control samples (dashed line).}
\label{plot_cf_simul}
\end{figure}

Assuming that our control sample is complete, we compute a cumulative 
distribution biased by the completeness 
function of Fig. \ref{plot_completude}. Fig. \ref{plot_cf_simul}
shows the result where it appears
 that the Malmquist bias in our sample, spread over a large distance range,
 can explain part of the apparent enhancement of molecular gas in the HGGs. 
However the control sample, assumed to be complete, is not
likely to be so for the low values of $\log(\mbox{\mhtwo / \lb}) )$ where 
the samples are the most different.
Similarly the $\log(\mbox{\mdust/\lb})$ distribution is affected by the 
Malmquist bias (Verter, 1993), but the point is that the cumulative 
distribution is lower than the control distribution on
the whole range of variation up to higher values, leading to a suggestion 
of a real enhancement of the dust material in the HCGs. The close correlation 
between dust and molecular gas suggests that the \htwo\ content
is really enhanced in the HCGs. {\em As it has been shown previously, that 
enhancement is only significant for the most compact groups in their merging 
phase, confusing somewhat the question of molecular gas enhancement in
the whole HCG sample.}

\subsection{High SFE due to artificially low gas content}

What can also be seen in Fig. \ref{fir-H2-SFE-HI}(a) is the large
dispersion of the SFE as defined by the \lfir/\mhtwo\ ratio: a large
number of galaxies are deficient in CO emission, leading to a depressed
\mhtwo/\lb\ ratio and large SFE. This large dispersion is mainly due to
the dwarf and elliptical samples, but also to small galaxies in the
control sample; we have checked that the objects with high SFE at
moderate \lfir/\lb\ have a lower \lb\ than average. This phenomenon
disappears when the total gas content is considered instead of the mere
H$_2$ content, as shown in fig \ref{fir-H2-SFE-HI}(b) and (d).  It is
striking that the total normalised gas content is almost a constant,
independent of \lfir/\lb. There is one exception for the cluster
population where the stripping of the neutral and molecular content is
at play.  We find a deficiency of the molecular content in these
galaxies, which seems related to a lower \lfir\ luminosity (Horellou et
al. 1995).  All that suggests the importance of \hi\ in star formation
as a source of fueling, through the conversion \htwo $\leftrightarrow$
\hI,  and the higher reliability of the star formation indicator
\lfir/(\mhtwo+\mhi). A least square fit yields the relation
$\lfir/(\mhtwo+\mhi) \simeq 2.4(\frac{\lfir}{\lb})^{0.69 \pm 0.10}$ 
for all the samples gathered.

\begin{figure*}
\centering
\psfig{width=14cm,height=10cm,file=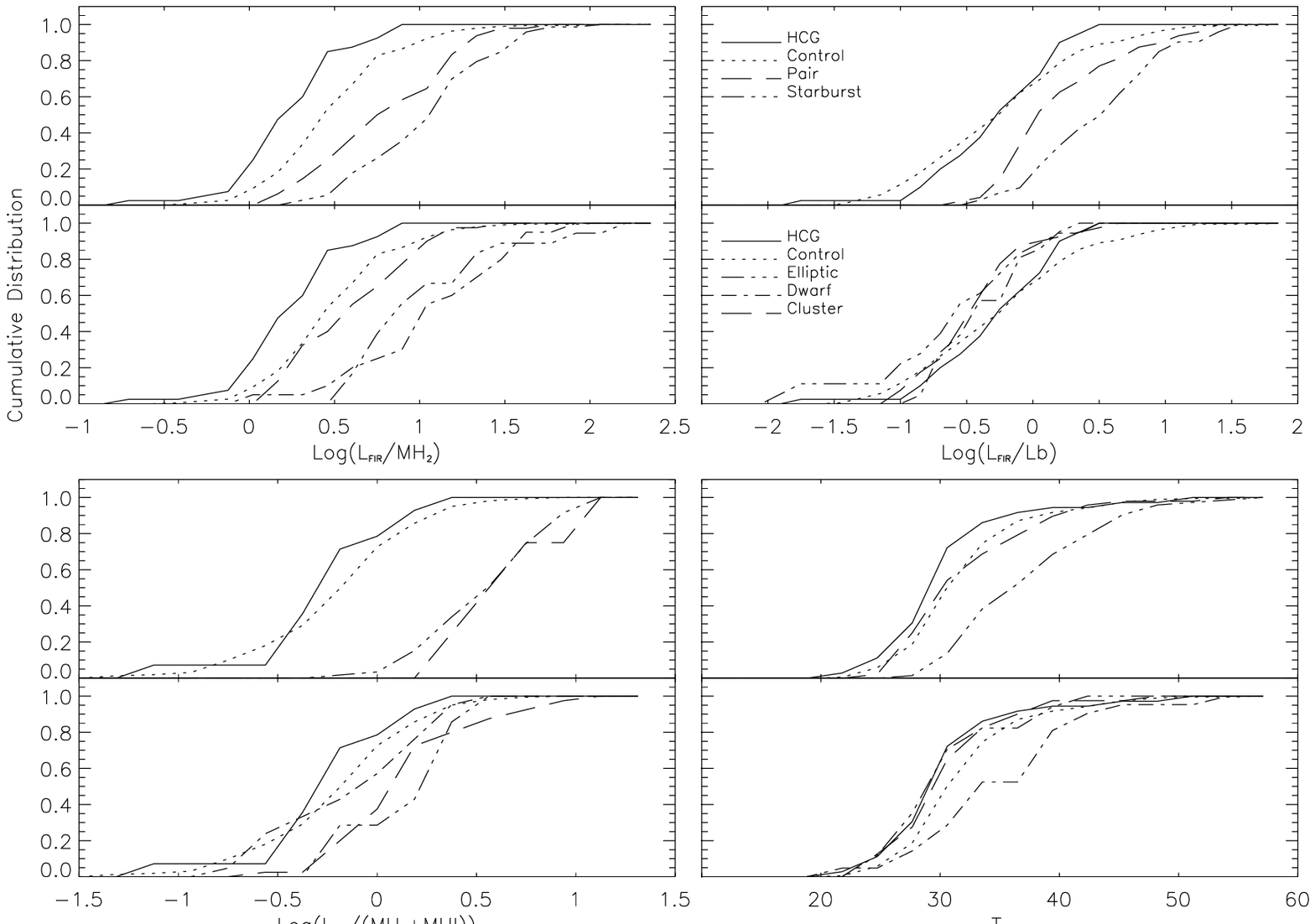}
\caption{Cumulative distribution of the different samples for the FIR-related
values.}
\label{df_fir}
\end{figure*}

The good correlation between the normalised \mhtwo\ and \lfir\ could be
in a large part due to the dependence of both quantities on the
metallicity and temperature of the interstellar medium. It is now well
established that the CO to H$_2$ conversion ratio is strongly dependent
on metallicity (e.g. Rubio et al 1993), as well as the dust-to-gas
ratio (and therefore the FIR luminosity). As shown on Fig.
\ref{df_ratio}, all the cumulative distributions for the molecular gas-to-dust
ratio are highly correlated, (\chitwo: all, except dwarf and elliptic samples, $<$
6. (probability $>$ 0.54)). While the dependence of \lfir\ on temperature is direct, that of
the CO emission is more complex. Low brightness temperatures of the CO
lines are obtained either for cold gas, or diffuse gas; the rotational
levels of the molecule are excited by collision, and diffuse molecular
clouds are generally sub-thermal. The use of a standard CO to H$_2$
conversion ratio is then problematic. For thermalised dense gas
however, the CO emission is directly proportional to the gas
temperature. The metal abundance and the gas recycling are closely
related to the IMF and evolution of the MF.  High mass stars (M$>
5\msol$) are active on short time scales ($<10^8$ yrs) whereas low mass
stars have an influence on much longer time scales (Vigroux et al.
1996). Thus the CO abundance is very dependent on the star formation
rate, and the CO/\htwo\ conversion ratio could be very variable,
particularly for galaxies with recent star formation episodes (e.g.
Casoli et al 1992, Henkel \& Mauersberger 1993).

\subsection {Far-infrared and star-forming activity}

Given the enhancement in CO emission detected in HCG with respect to a
control sample and the strong correlation between the radio continuum,
CO and the FIR for HCG spirals it would appear that the lack of
enhancement of the total FIR emission in HCG galaxies is due to lack of
spatial resolution of IRAS measurements. This is particularly
important  if the FIR is mainly enhanced in the central regions of the
galaxies.  ISO observations might allow to check this assertion.
 
  The consequence of an enhanced \mhtwo\ without FIR enhancement is
a lower star forming efficiency for HCG, as displayed in Fig. 
\ref{df_fir}. This property might appear surprising, but disappears
when the total gas content is taken into account (this result should
be taken with caution, since only 14 galaxies have a well-defined
\hi\ content in our HCG sample).

\begin{figure*}
\centering
\caption{ Comparison between the HCG sample and other CO-detected samples:
(a)-- \lfir/\mhtwo\ versus \tdust. 
(b)-- \lfir/(\mhtwo\ + M(HI)) versus \tdust. 
(c)-- \mhtwo/\lb versus \tdust.
(d)-- (\mhtwo\ +\mhi)/\lb versus \tdust.}
\label{survey_td}
\psfig{width=18cm,height=14cm,file=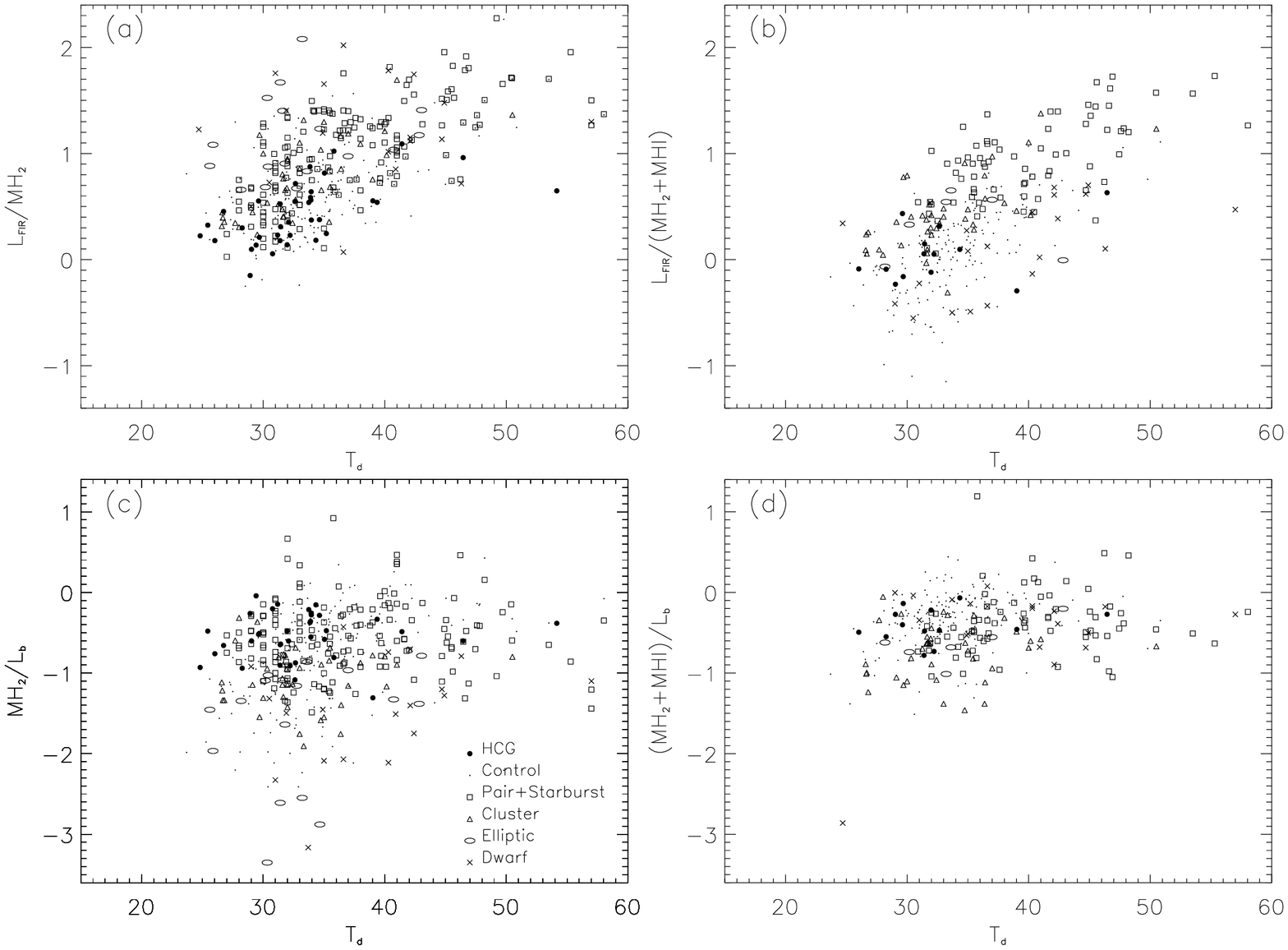,angle=90}
\end{figure*}
\label{Td}

  We have tested the correlations of the star formation indicators, with
and without account of the atomic phase, with the dust temperature, in
Fig. \ref{survey_td}.  Both quantities correlate well with \tdust: for all
samples together  we find  a relationship flatter than  Young et al. (1989),
i.e.  \lfir/\mhtwo $\propto$  $\tdust^{3.8\pm 0.7}$, but with differences 
and large dispersions
among categories:  for the HCG sample alone \lfir /\mhtwo $\propto$
$\tdust^{2.5 \pm 2.5}$ and for the starburst sample \lfir/\mhtwo $\propto$
$\tdust^{3.2 \pm 1.8}$.  Sage (1993) pointed out that a single dust temperature
is an ``average'' of the cold dust and warm dust associated
respectively with the quiescent molecular clouds and the clouds with
massive star forming ones.  Devereux \& Young (1990) emphasized that
high mass ($>$ 6 \msol) O and B stars are responsible for high
\lfir\ ($> 10^9$ \lsol) and H$_\alpha$ luminosities, advocating a
two-component model: one dust component heated by high mass stars
(\tdust $\sim$ 50-60 K) and the other heated by the interstellar
radiation field (\tdust $\sim$ 16-20 K).  The low mean temperature of
the HCG sample (\tdust $\approx$ 33 K) suggests that the
\lfir\ luminosity is coming from an important quiescent molecular
phase, together with the ``cirrus'' phase associated with diffuse
atomic hydrogen.  It could explain  why the {\em global} \lfir\
luminosity does not fit with a high star formation population, without
excluding star formation towards the center, as revealed by radio
continuum.  In a recent study, Lisenfeld et al. (1996) find the same
\lfir/\ltwogiga\ ratio for starburst interacting and normal galaxies.
From this, we could infer that the enhanced radio continuum emission
from the center of HCG galaxies should be accompanied with an enhanced
central FIR luminosity. The constancy of the \lfir/\ltwogiga\ ratio has
been interpreted as being due to a strong and fast ($10^7$ yrs)
increase of the magnetic field at the beginning of the starburst,
together with a time-scale of variation of the star-formation rate
longer than some $10^7$ yrs.

\subsection{Fate of the HCG}

The enhancement of the \htwo\ content in the most compact groups
suggests that tidal interactions in HCG are efficient in  driving the
gas inwards.  This is also a confirmation that at least some groups are
actually compact and not only projections along the line of sight.
These very compact groups should merge through dynamical friction on a
short time scale (a few $10^8$ yrs, cf Barnes 1989).  A conclusion from
the present work is that the most compact of these groups have
concentrated an important amount of molecular gas without initiating
yet important star formation. However they must correspond to a short
duration phase just before merging and enhanced star-formation.  It is
tempting to identify the next phase of this process to the Ultra
Luminous Infra Red Galaxies (ULIRGs).  The latter have their infrared
luminosity powered by massive star formation (Lutz et al. 1996) with an
important consumption of molecular gas.  Sanders et al (1988) have
shown that many ULIRGs are interacting/merging galaxies; recently
Clements \& Baker (1996) extended that to the vast majority of the
ULIRGs sample.  Some of these systems, with luminous masses ranging up
to few $10^{11} \msol$, should represent the remnant of some compact
group which has undertaken multiple mergers on a very short time-scale
($\sim 10^8$ yrs). The FIR luminosity of the final ULIRGs should be a
function of the spiral fraction in the parent group, the most powerful
ULIRGs being the result of the merging of typically four gas-rich
spirals. As pointed out earlier the spatial distribution of
interstellar matter within galaxies will play an important role in the
rate of fuelling of starbursts during interactions. 

\begin{figure}

 But does the expected rate of ULIRGs formation via CG merging match the
presently observed frequency of these objects?  The answer is very
uncertain, since the actual fraction of very compact groups, on the
point of merging, is not known. This fraction cannot be close to 1,
since there would be too large a discrepancy between the expected and
observed number of ultra luminous galaxies and their remnants (e.g.
Williams \& Rood 1987, Sulentic \& Raba\c{c}a 1994).  Simulations of
galaxy formation and large-scale structures evolution have suggested
that some of the CGs could be filaments of galaxies seen end-on
(Hernquist et al. 1995). This idea has been studied further by Pildis et
al. (1996): if it is true that galaxies that will form a compact group
spend a large fraction of their time first in a filament, this filament
will appear as a CG in projection only for less than 20\% of cases.
Then the galaxies will fall into a real CG, and this phase corresponds
to at least 30\% of their lifetime. Although these figures are model
dependent, they suggest that the majority of HCG in the sky are
physically compact groups.  The time-scale of merging can then depend
highly on initial conditions, and in particular on the elliptical
fraction, which may alleviate the over-merging problem (e.g. Governato
et al. 1991, Garcia-Gomez et al. 1996).

\centering
\psfig{width=8cm,height=6cm,file=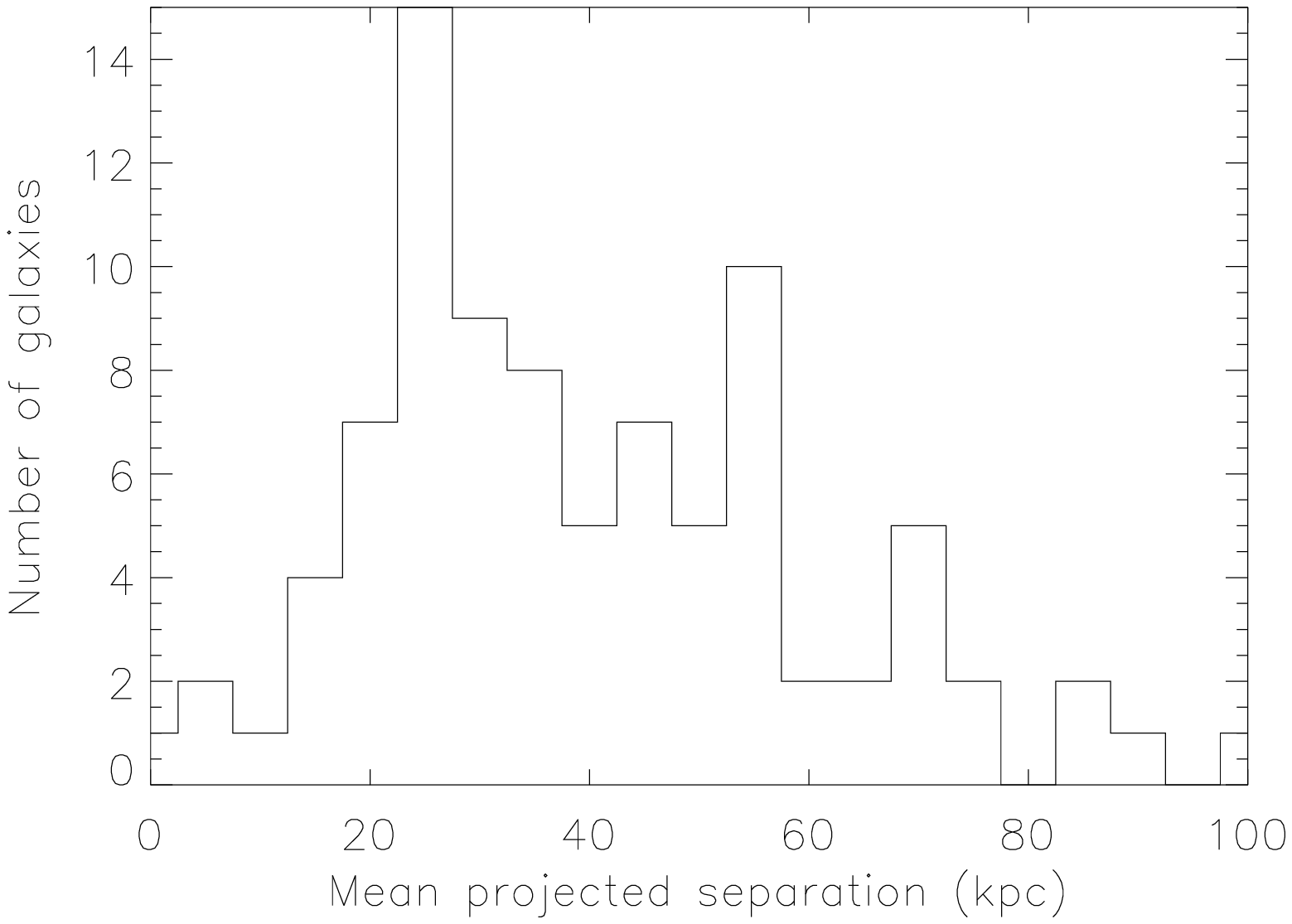}
\caption{Histogram of the mean projected separation for the whole HCG sample, 
limited to 100 kpc, from Hickson et al. (1992).}
\label{histo_sepa}
\end{figure}

The threshold in mean galaxy separation for the enhancement of \htwo\
suggests that it corresponds to the last stage of the life of the
compact group, when each galaxy is undertaking frequent and strong
tidal effects. In Fig. \ref{histo_sepa} we plot the histogram of the
mean projected separation of the whole sample of HCGs from Hickson et
al. (1992): there is a significant cut-off at short separation at
approximatively 20-30 kpc, which must correspond to a short life time
for very compact configurations. We suggest that this cut-off
corresponds to the acceleration of the merging process, and at the same
time to significant inward gas flows, that account for our
observations.  Dynamically, such an exponential acceleration of the
collapse is predicted by simulations and analytic models of satellite
decay through dynamical friction (see for example Leeuwin \& Combes
1997). This rapid acceleration has been interpreted through the
excitation of numerous high-order resonances for a satellite at about
twice the primary radius (e.g. Tremaine \& Weinberg 1984)
.

\section{Conclusion}

From our survey of CO emission in 70 galaxies belonging to 45
Hickson compact groups, we have detected 57 objects. We find in average
that the gas and dust contents \mhtwo/\lb\ and \mdust/\lb\ show evidence
of enhancement with respect to our control sample. 
This result however is somewhat weakened due to the 
Malmquist bias in our sample. For the most compact groups the
enhancement is more clear. On the contrary, the global
far-infrared  flux does not appear to be enhanced with respect to the
control sample. The FIR and \tdust\ distribution indicates that the FIR
luminosity is coming essentially from a cold dust component heated by
the interstellar radiation field.
From the general correlation between FIR and radio contiuum power we
suggest that only the very centers of some groups are experiencing star
formation and are sites of  enhanced FIR emission.  IRAS spatial
resolution is not sufficient to show this directly.  Statistical tests
show that the HCG gas and dust contents are closer to that of pair and
starburst galaxies, revealing the efficiency of tidal interactions in
driving the gas inwards in compact group galaxies.  We find a stronger
\htwo\ enhancement for the CGs having a short mean separation ($< 30$
kpc). We suggest that these most compact, high-\htwo\ content groups,
may be in a final merging phase, just before the starburst phase, that
will lead them in a very short time-scale to the ULIRGs category.\\

The comparison of the various samples suggest that the {\em total} gas
content (\htwo+\hi) should be taken into account to estimate the star
formation efficiency. The corresponding SFE indicator,
\lfir/(\mhtwo\ +\mhi), should be more reliable, and allow us to avoid
some systematic effects depending on metallicity and temperature.

\appendix
\section{Appendix: H$_2$ mass determination}

\begin{figure}
\label{fig_K}
\caption{Correction factor K in the case of an exponential (solid line) 
and an uniform (dot line)  source distribution}
\psfig{width=8cm,height=6cm,file=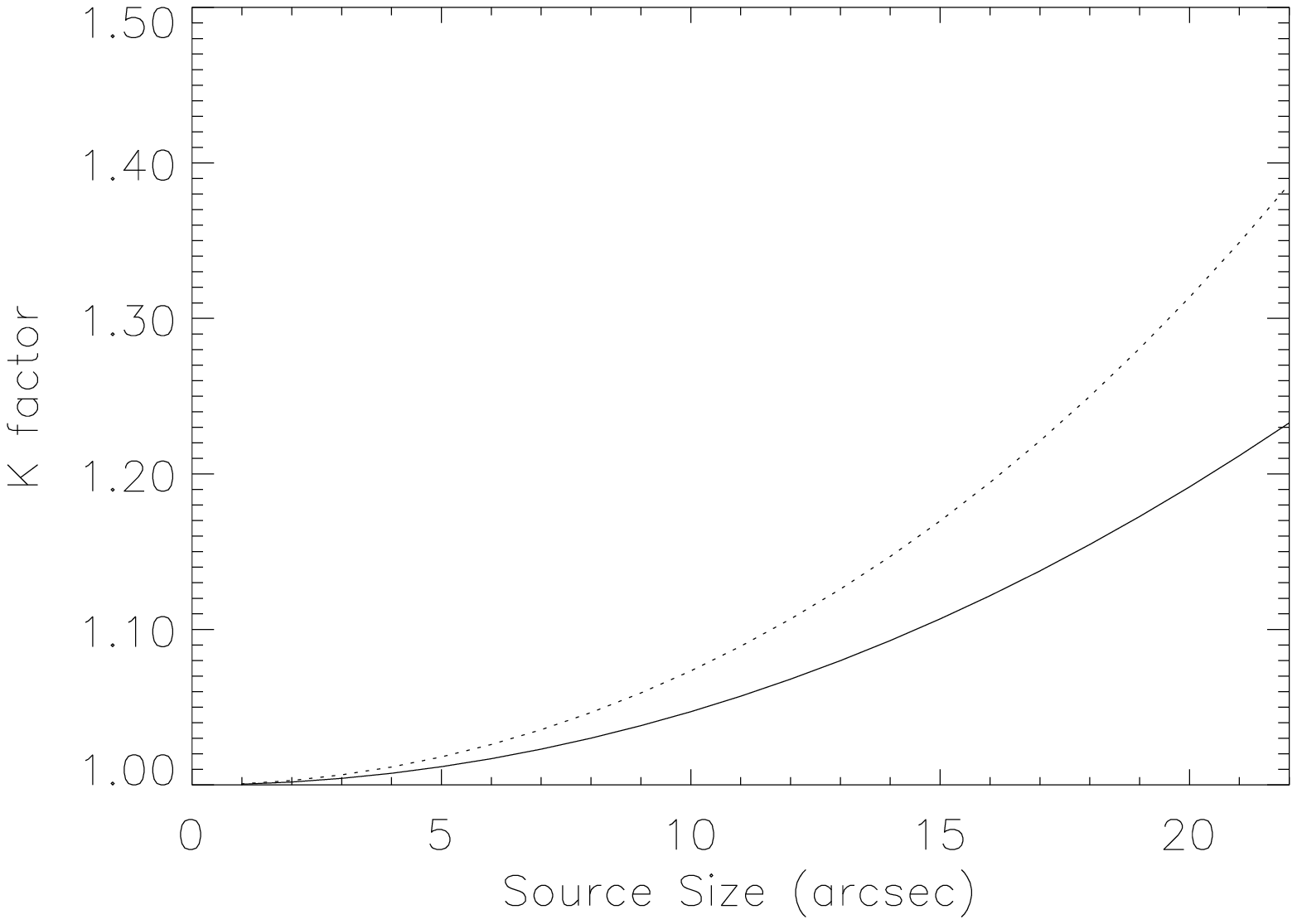}
\end{figure}

We follow Gordon et al. (1992) to derive H$_2$ masses from \CO{1}{0}
line observations. Our temperature unit is expressed in \tastar antenna
temperature scale which is corrected for atmospheric attenuation and
rear sidelobes. The radiation temperature T$_R$ of the extragalactic
source is then:

\begin{equation}
\label{equa_Tr}
T_R=\frac{4}{\pi}(\frac{\lambda}{D})^2\frac{K}{\eta _A} \frac{T_A}{\Omega_S}
\end{equation}

where $\lambda$ is the observed wavelength (2.6 mm), D is the IRAM
radio telescope diameter (30 m), K is the correction factor for the
coupling of the source with the beam, $\eta_A$ is the apperture
efficiency (0.55) at 115 GHz, T$_A$ is an antenna temperature which is
$F_{eff}$T$_A^*$ in the IRAM convention, explicitely
T$_A$=0.92T$_A^*$, and $\Omega_S$ is the source size. Without taking
into account cosmological correction, because of low redshift, column
density of molecular hydrogen it written down as
  
\begin{equation}
     N(H_2)=2.3\times10^{20}\int_{line} T_R dv  \mbox{ (mol.cm$^{-2}$)}
\end{equation}
where $dv$ is the velocity interval .
In equation \ref{equa_Tr} K$\equiv \Omega_s / \Omega_{\Sigma}$ is the factor 
which corrects the measured antenna temperature
for the weighting of the source distribution by the large antenna beam in 
case of a smaller source. We have defined the source solid angle

\begin{equation}
\Omega_S\equiv \int_{source} \phi (\theta , \psi ) d\Omega
\end{equation}
where $\phi (\theta , \psi$) is the normalized source brightness distribution
 function. The beam-weighted source solid angle is

\begin{equation}
\Omega_{\Sigma}\equiv \int_{source} \phi (\theta , \psi ) f( \theta, \psi ) 
d\Omega
\end{equation}
where $f( \theta, \psi )$ denotes the normalized antenna power pattern
(Baars, 1973). Experiments have shown that we can approximate $f$ by a
gaussian beam. As mentionned in section 3.3 an exponential law of scale
length h=D$_B$/10 is taken to model the source distribution function.
As long as the source size is smaller than the beam size we have

\begin{equation}
\label{equa_K}
K=\frac{\int^{\theta_s /2}_0 sin(\theta ) e^{-\frac{ 10 \theta }{\theta_s}} 
d\theta}{\int^
{\theta_s /2}_0 sin(\theta ) e^{-\frac{ 10 \theta }{\theta_s}-\mbox{ln}(2)(
\frac{2 \theta}{\theta_b})^2} d\theta}
\end{equation}
where $\theta_b$ and $\theta_s$ are the beam and the source sizes. We
present in Fig. \ref{fig_K} the plot of K in the case of exponential
and uniform source distributions, for the IRAM-30m beam at 115 GHz (22
\arcsec). In the case of a source larger than the beam size it is more
difficult to compute the coupling of  the  beam to the source region
since contributions from error beam can be quite significant to the
resulting spectra. A simple representation of the overall beam,
including the error beam, is not available to correct that coupling.
Since in our sample, galaxies are at most a few beam sizes large we
used equation \ref{equa_K} integrated on the beam size to derive the K
factor for galaxies with larger optical diameters.\\

If I$_{CO}$ is the velocity-integrated temperature in T$_A^*$ scale and
given the IRAM-30m  parameters, the total mass of H$_2$ is then given
by

\begin{equation}           
M(H_2)=5.86\times10^4D^2KI_{CO} \mbox{ (\msol)}
\end{equation}
with the distance D in Mpc.

\begin{acknowledgements}
  We are very grateful to the staff at Pico Veleta for their help during
these observations and especially to R. Moreno and R. Frapolli. We also
thank the anonymous referee whose suggestions greatly improved this paper.
 This work largely benefitted from the LEDA, NED and
SIMBAD data bases. This investigation was also supported by a grant
from the Natural Sciences and Engineering Research Council of Canada to T.K.M.
    
\end{acknowledgements}

\end{document}